\documentclass[aps,pra,10pt,twocolumn,superscriptaddress,amssymb,amsmath, showpacs,balancelastpage, nofootinbib]{revtex4-1}
\usepackage{CJKutf8}

\usepackage{amsthm}

\usepackage[utf8]{inputenc}
\usepackage{amsmath,amssymb,amsfonts,stmaryrd}
\usepackage{physics}
\usepackage{dsfont}
\usepackage{graphicx}
\usepackage{xcolor}
\usepackage{graphicx}
\usepackage{cancel}
\usepackage{natbib}
\usepackage{color}
\usepackage{tikz}
\usepackage{mathrsfs}
\usepackage{relsize}
\usepackage{multirow}
\usepackage[linktocpage]{hyperref}
\usepackage[all]{xy}
\usepackage[normalem]{ulem}
\hypersetup{colorlinks=true,citecolor=blue,linkcolor=blue, urlcolor=blue, breaklinks=true}

\usepackage{float}

\usepackage{bm} %allows for bold math type
\usepackage{subfigure} %allows for 2x2 figures
\usepackage{environ}
\usepackage{url}
\usepackage[margin=0.75in]{geometry}

\usepackage{mathtools}

\newcommand{\Z}{{\mathbb Z}}

\def\U{\mathrm{U}(1)}
\newcommand{\nfs}{{N_\mathrm{FS}}}
\def\TT{\mathcal{T}}

\NewEnviron{eqs}{%
\begin{equation}\begin{split}
    \BODY
\end{split}\end{equation}}

\definecolor{dkgreen}{rgb}{0,0.5,0}

\theoremstyle{definition}

\theoremstyle{remark}

\begin{document}

\begin{CJK*}{UTF8}{bsmi}

\title{Higher Hall conductivity from a single wave function:
\texorpdfstring{\\}{}
Obstructions to symmetry-preserving gapped edge of (2+1)D topological order}

\author{Ryohei Kobayashi}
\email[E-mail: ]{ryok@ias.edu}
\affiliation{Department of Physics, Condensed Matter Theory Center, and Joint Quantum Institute, University of Maryland, College Park, Maryland 20742, USA}

\affiliation{School of Natural Sciences, Institute for Advanced Study, Princeton, NJ 08540, USA}

\author{Taige Wang}
\affiliation{Department of Physics, University of California, Berkeley, California 94720 USA}
\affiliation{Material Science Division, Lawrence Berkeley National Laboratory, Berkeley, California 94720, USA}
\affiliation{Kavli Institute for Theoretical Physics, University of California, Santa Barbara, California 93106, USA}

\author{\mbox{Tomohiro Soejima (副島智大)}}
\affiliation{Department of Physics, Harvard University, Cambridge, MA 02138, USA}

\author{Roger S. K. Mong (蒙紹璣)}
\affiliation{Department of Physics and Astronomy,
University of Pittsburgh, Pittsburgh, PA 15260, USA}

\author{Shinsei Ryu}
\affiliation{Department of Physics, Princeton University, Princeton, New Jersey, 08544, USA}

\date{\today}
\begin{abstract}
A (2+1)D topological ordered phase with U(1) symmetry may or may not have a symmetric gapped edge state, even if both thermal and electric Hall conductivity are vanishing. It is recently discovered that there are ``higher'' versions of Hall conductivity valid for fermionic fractional quantum Hall (FQH) states, which obstructs symmetry-preserving gapped edge state beyond thermal and electric Hall conductivity. In this paper, we show that one can extract higher Hall conductivity from a single wave function of an FQH state, by evaluating the expectation value of the ``partial rotation'' unitary which is a combination of partial spatial rotation and a U(1) phase rotation. This result is verified numerically with the fermionic Laughlin state with $\nu=1/3$, $1/5$, as well as the non-Abelian Moore-Read state.
Together with topological entanglement entropy, we prove that the expectation values of the partial rotation completely determines if a bosonic/fermionic Abelian topological order with U(1) symmetry has a symmetry-preserving gappable edge state or not. 
We also show that thermal and electric Hall conductivity of Abelian topological order can be extracted by partial rotations.
Even in non-Abelian FQH states, partial rotation provides the Lieb-Schultz-Mattis type theorem constraining the low-energy spectrum of the bulk-boundary system. The generalization of higher Hall conductivity to the case with Lie group symmetry is also presented. 
 \end{abstract}

\maketitle

\end{CJK*}

%\tableofcontents

%\bigskip

%%%%%%%%%%%%%%%%%%%%%%%%%%%%%%%%%%%%%%%%%%%%%%%%%%%%%%%%%%%%%%%%%%%%%%%%%%%%%

\section{Introduction}

One of the most striking properties in topological phases of matter is the appearance of non-trivial edge states
\cite{PhysRevB.25.2185,
PhysRevLett.64.2206,Stone:1990iw,
Frohlich:1990xz}.
The most well-known example is the Integer Quantum Hall (IQH) effect, where non-trivial bulk Chern number leads to the presence of gapless edge state with U(1) symmetry~\cite{hatsugai1993}, which can be probed by the Hall conductance $\sigma_H$. 
Even in the absence of U(1) symmetry, systems with nonzero chiral central charge $c_-$, which signals nonzero thermal Hall conductivity, have a gapless edge state~\cite{KaneFisher}.   
The invariants including thermal or electric Hall conductivity can be characterized through the bulk effective theory, as well as extracted from a microscopic wavefunction, through various entanglement measures such as modular commutator, through randomized measurement, and so on~\cite{Kitaevanyons, Mitchell2018amorphus, Qi2012momentumpolarization, FQHEDMRG, Kim2022cminus, Kim2022modular, Zou2022modular, Fan2022cminus, Fan2022QHE, Shiozaki2018antiunitary, Dehghani_2021, Cian_2021,liang2024extracting, liang2024operator, kobayashi2024universal}.

When the bulk hosts a topologically ordered state, the edge state is sometimes enforced to be gapless even if electric and thermal Hall conductivity are both vanishing. This happens when one cannot condense a sufficient set of anyons at the boundary enough to gap out the edge theory~\cite{Kapustin:2010hk, Levin2013edge}. When the U(1) symmetry is taken into account, the anyon condensation on the boundary further has to respect the global symmetry to host a symmetric edge state.

Recently, it is discovered that a set of quantities called the higher central charge~\cite{Ng2018higher, Ng2020higher, kaidi2021higher} and the higher Hall conductivity~\cite{Kobayashi2022FQH} partially give obstructions to having a gapped edge state beyond thermal and electric Hall conductivity. 
Concretely, for bosonic topological order in (2+1)D, having non-trivial higher central charge $\zeta_n$ for some integer $n$ prohibits us from gapping out the edge state even if $c_-=0$. Similarly, let us consider fermionic topological order with $\U^f$ symmetry in (2+1)D, where $\U^f$ denotes $\U$ symmetry that charges a local fermion. This is a global symmetry of fermionic fractional quantum Hall (FQH) states. In that case, having non-trivial higher Hall conductivity again prohibits us from gapping the edge state while respecting $\U^f$ symmetry, even if $c_-=\sigma_H=0$.

While both the higher central charge and higher Hall conductivity are characterized through effective topological field theory, it is an important question to ask if these quantities allow microscopic characterizations. In the prior work by the authors \cite{Kobayashi2024hcc}, it was found that the higher central charge can be extracted from a single wavefunction, by evaluating the expectation value of a unitary called the partial rotation. The extracted higher central charge then allows us to constrain the low-energy spectrum of the bulk-boundary system, e.g., completely determine if the given wavefunction bosonic of Abelian topological order has a gappable edge state or not.
In this work, we study microscopic characterization of the higher Hall conductivity in the presence of $\U^f$ symmetry, and explore its physical consequences including constraints on the low energy spectrum of edge states enforced by symmetry.

\subsection{Summary of results}
Here we summarize the main results of this paper. One of the main results is that one can extract the higher Hall conductivity $\{\zeta_n^H\}$, which is a set of invariants labeled by an integer $n$ satisfying certain conditions. %which are coprime to $\nfs$---i.e.\, $\gcd(n,\nfs)=1$---where $\nfs$ is an integer called Frobenius-Schur exponent.

\begin{figure}[htbp]
    \centering
    \includegraphics[width = 0.3\textwidth]{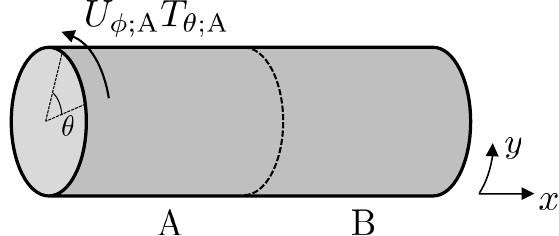}
    \caption{The invariant $\mathcal{T}_a\left(\theta,\phi\right)$ is defined as an expectation value of a unitary $U_{\phi,\mathrm{A}}T_{\theta,\mathrm{A}}$ acting within the A subsystem of the cylinder.}
    \label{fig:rotation}    
\end{figure}

Let us consider a single wave function of a fermionic FQH state on a cylinder $\ket{\Psi_a}$ with a fixed topological sector, which is labeled by a simple anyon of the bulk topological order $a$.
We take the bipartition of the cylinder into the subsystem A and B, see Fig.~\ref{fig:rotation}.
We then compute the expectation value of the ``partial rotation'' unitary, which is the product of the partial spatial rotation $T_{\theta,\mathrm{A}}$ associated with the partial $\U$ transformation $U_{\phi,\mathrm{A}}$ on the wave function,
\begin{align}    \mathcal{T}_a\left(\theta,\phi\right):=\bra{\Psi_a}U_{\phi,\mathrm{A}}T_{\theta,\mathrm{A}} \ket{\Psi_a}.
\end{align}
In a slight abuse of notation, we also refer to the expectation values as partial rotation when it is clear from context. The case of $\phi=0$ reduces to the partial rotation operator introduced in Ref.~\onlinecite{Kobayashi2024hcc}.
We show that a nontrivial phase of the above quantity $\mathcal{T}_a\left(\theta,\phi\right)$ with specific rotation angle is 
an obstruction to having a gapped edge state while preserving $\U^f$ symmetry. That is, for integers $n$ satisfying the condition $\gcd(n,\nfs)=1$ with $\nfs$ an integer called the Frobenius-Schur exponent, we demonstrate that
\begin{align}
    \mathcal{T}_1\left(\frac{2\pi}{n},\frac{\pi}{n}\right)  & \propto \exp\biggl[ \frac{2\pi i}{n} \bigg(\frac{\sigma_H}{8}-\Big(\frac{n^2+2}{24}\Big)c_-\bigg) \biggr] \zeta_n^H,
    \end{align}
    where $\propto$ in this paper always means being proportional up to a positive real number.
    This quantity is the combination of $\sigma_H$, $c_- $, and $\zeta_n^H$. The latter gives an obstruction to the presence of a 
symmetric gapped edge state beyond $\sigma_H$, $c_-$.

Since the non-trivial higher Hall conductivity prohibits the symmetry-preserving gapped edge state of the given wave function, the expectation values of the partial rotation can be used to determine or constrain the low-energy spectrum of the edge state enforced by symmetry. In particular, when the topological order is Abelian, one can completely determine if the given wavefunction admits a symmetry-preserving gapped edge or not. Our result 
is supported by an analytical computation in terms of conformal field theory (CFT), as well as numerics on 
fermionic Laughlin states and the
non-Abelian Moore-Read FQH state.

Next, let us describe our result valid for Abelian topological order. We will first see that both electric and thermal Hall conductivity can be extracted by partial rotation and topological entanglement entropy of a single wavefunction.\footnote{Throughout the paper, we assume that topological entanglement entropy gives total quantum dimension $\mathcal{D}$ of the topological order, i.e., there is no spurious contributions as discussed in \cite{Cano2015Interaction, Zou2016spurious, Williamson2019spurious, Kim2023bound, Levin2024inequality}. This assumption is valid when the entanglement spectrum is given by the edge CFT, which is also used to calculate the partial rotations in this paper.} This is valid in either fermionic or bosonic Abelian phases, and we numerically verify the result on a bosonic Laughlin states.
Even when $\sigma_H=c_-=0$, we show that once the topological entanglement entropy of the given wave function is known, the partial rotations completely determine
if a given wave function for Abelian topological order admits a gapped edge state preserving $\U$ symmetry or not. This statement is again valid
for either bosonic or fermionic Abelian phases.

While partial rotation does not span complete obstructions to symmetric gapped edge state for non-Abelian topological order, it is still useful for constraining the low-energy spectrum of the bulk-boundary system. For (2+1)D fermionic non-Abelian topological order with $\U^f$ symmetry,  we show that the suitable set of expectation values of partial rotation enforces the bulk-boundary spectrum to be 1: gapless, 2: symmetry broken, or 3: symmetry-preserving gapped edge state with non-trivial lower bound on the number of anyons in the bulk.
The third possibility requires an extensive number of bulk ground state degeneracy to realize the symmetry-preserving gapped boundary. This is reminiscent of the Lieb-Schultz-Mattis type theorem~\cite{Lieb1961, Oshikawa, Hastings}, which constrains the low-energy spectrum of a
quantum many-body system with given input of symmetry action on the state.

Finally, while higher Hall conductivity has been defined only for topological order with U(1) symmetry, we propose its generalization to the case with Lie group symmetry. In that case, higher Hall conductivity again gives obstructions to gapped edge state preserving the Lie group symmetry, and can be extracted from partial rotation.

This paper is organized as follows. In Sec.~\ref{sec:property}, we review the definitions and properties of higher Hall conductivity and higher central charge. In Sec.~\ref{sec:partial}, we show that the higher Hall conductivity can be extracted from a single wave function through partial rotation, and also that $c_-$ and $\sigma_H$ of Abelian topological order can be extracted through partial rotation and topological entanglement entropy. In Sec.~\ref{sec:numerics}, we provide numerical results from DMRG for FQH states 
on a cylinder, and verify our analytical results. In Sec.~\ref{sec:complete}, we explicitly show that one can completely determine the edge gappability of Abelian topological order with U(1) symmetry. In Sec.~\ref{sec:lsm}, we show the Lieb-Schultz-Mattis type theorem for bulk-boundary systems of fermionic non-Abelian topological order with $\U^f$ symmetry. In Sec.~\ref{sec:lie}, we describe a generalization of the higher Hall conductivity for the phases with Lie group symmetry. In Sec.~\ref{sec:discussion}, we conclude our paper with possible future work.

\section{Definition and properties of higher Hall conductivity}
\label{sec:property}

\subsection{Higher central charge of bosonic topological order}
For completeness, let us first review higher central charge of (2+1)D bosonic topological order that obstructs the presence of a
gapped boundary. 
Given a modular category $\mathcal{C}$, $\nfs$ is the Frobenius-Schur exponent defined as the smallest positive integer satisfying $\theta_a^\nfs=1$ for all anyons $a\in\mathcal{C}$.

The higher central charge $\zeta_n$ is labeled by a positive integer $n$, which is defined as
\begin{align}
     \zeta_n :=\frac{\sum_{a\in\mathcal{C}}d_a^2\theta_a^n}{\bigl| \sum_{a\in\mathcal{C}}d_a^2\theta_a^n \bigr|}
     \label{eq:hcc}
\end{align}
where $d_a$, $\theta_a$ are the quantum dimension and the topological twist of the anyon $a$, respectively.
The sum runs over all the anyons $a$ of the bosonic phase, which are simple objects of the modular tensor category $\mathcal{C}$.

The trivial higher central charge $\zeta_n=1$ for all $n$ such that $\gcd(n,\nfs)=1$ gives necessary conditions to gapped boundary of bosonic topological order, which are generally independent of the vanishing chiral central charge condition $c_-=0$~\cite{Ng2018higher}. This implies that the set of quantities $\{\zeta_n\}$ prohibits the existence of the gapped edge state even if $c_-$ is vanishing. 
In the special case with $n=1$, $\zeta_1$ just gives the Gauss-Milgram formula for $c_-$ mod 8; $\zeta_1= e^{2\pi i c_-/8}$. In this sense, $\zeta_n$ with $n>1$ can naturally be regarded as a ``higher'' version of $c_-$, hence the name of higher central charge.

When the topological order is Abelian, $\zeta_n=1$ for all $n$ satisfying $\gcd\bigl( n,\frac{\nfs}{\gcd(n,\nfs )} \bigr)=1$ gives necessary and sufficient conditions for the gapped boundary~\cite{kaidi2021higher},
so $\{\zeta_n\}$ completely determines the gappability of the edge state of Abelian topological order.%~\footnote{Though we defined $\zeta_n$ only when $\gcd(n,\nfs)=1$ in Eq.~\eqref{eq:hcc}, we can define $\zeta_n$ by the same formula as~\eqref{eq:hcc} for $n$ satisfying $\gcd(n,\frac{\nfs}{\gcd(n,\nfs )})=1$ in the Abelian case. } 

\subsection{Higher Hall conductivity of fermionic FQH states}

Here we review the definition and properties of the higher Hall conductivity, which is valid for the (2+1)D fermionic topological phase with $\U^f$ symmetry~\cite{Kobayashi2022FQH}.

For a super-modular category $\mathcal{C}$ that describes the fermionic topological order, we again define the Frobenius-Schur exponent $\nfs$, defined as the smallest positive integer satisfying $\theta_a^\nfs=1$ for all anyons $a\in\mathcal{C}$.
$\nfs$ must be an even integer since the fermion has topological spin $1/2$.

The higher Hall conductivity $\zeta_n^H$ is again labeled by positive integers $n$, which is defined as the following quantity computed from the data of anyons:
\begin{align}
    \zeta^H_n := 
    \dfrac{\sum_{a\in\mathcal{C}}e^{i\pi Q_a} d_a^2\theta_a^n}{\bigl| \sum_{a\in\mathcal{C}}e^{i\pi Q_a} d_a^2\theta_a^n \bigr|}
\end{align}
The sum runs over all the anyons $a$ of the fermionic phase, which are formally regarded as simple objects of the super-modular tensor category $\mathcal{C}$. $Q_a$ denotes the fractional $\U^f$ charge of the anyon $a$.

Importantly, $\zeta_n^H=1$ for all $n$ satisfying $\mathrm{gcd}(n,\nfs)=1$ gives necessary conditions to having a 
symmetry-preserving gapped edge state, generally independent of the most fundamental conditions $c_-=\sigma_H=0$~\cite{Kobayashi2022FQH}.
This implies that the set of quantities $\{\zeta^H_n\}$ prohibits the existence of the symmetry-preserving gapped edge states even if both $c_-$ and $\sigma_H$ are vanishing. 
In that sense, $\zeta^H_n$ is regarded as a ``higher'' version of thermal and electric Hall conductivity,
hence the name of higher Hall conductivity.

We note that $c_-$ and $\sigma_H$ are also computed modulo 1 from the data of anyons as~\cite{kobayashi202331d}
\begin{align}
    e^{-2\pi i c_-}
    &= (\sqrt{2}\mathcal{D})^8\cdot \frac{\sum_{a\in\mathcal{C}}e^{3i\pi Q_a} d_a^2\theta_a}{\left(\sum_{a\in\mathcal{C}}e^{i\pi Q_a} d_a^2\theta_a\right)^9}~,
    \label{eq:c-spinc}
    \\
    e^{-2\pi i {\sigma}_H}
    &= \frac{\sum_{a\in\mathcal{C}}e^{3i\pi Q_a} d_a^2\theta_a}{\sum_{a\in\mathcal{C}}e^{i\pi Q_a} d_a^2\theta_a}~,
    \label{eq:sigmaspinc}
\end{align}
where $\mathcal{D}$ is the total quantum dimension $\mathcal{D}^2 = \sum_a d_a^2$.

The higher Hall conductivity $\zeta_n^H$ cannot generally be expressed as a linear combination of $c_-,\sigma_H$.
Meanwhile, in the special case with $n=1$, we have the relation~\cite{Lapa2019}
\begin{align}
    \zeta_1^H = e^{\frac{2\pi i}{8}(c_--\sigma_H)}~.
\end{align}

\section{Extracting higher Hall conductivity from a single wavefunction}
\label{sec:partial}
\subsection{Partial rotation for the fermionic phase}
In this section, we show that higher Hall conductivity $\zeta_n^H$ can be extracted from a single wave function, by evaluating the expectation value of the partial spatial rotation followed by partial $\U^f$ symmetry transformation.

Let us consider a (2+1)D fermionic topological ordered state with $\U^f$ symmetry on a cylinder. The state on the cylinder is labeled by the anyon $a$, which corresponds to a quasiparticle obtained by shrinking the puncture at the end of the cylinder.
We denote the ground state $\ket{\Psi_a}$ on the cylinder with the anti-periodic boundary condition for $\Z_2^f$ fermion parity symmetry. In that case, the state is labeled by the anyon $a$ of the theory, which is the object of the super-modular category $\mathcal{C}$.

Let us take a bipartition of the cylinder into the two subsystems labeled by A and B. We write the translation operator for the A subsystem by the angle $\theta$ along the circumference as $T_{\theta;\mathrm{A}}$, and also the ``partial'' $\U^f$ transformation by the angle $\phi$ that solely acts on the A subsystem as $U_{\phi,\mathrm{A}}$, where $\phi=\pi$ corresponds to the fermion parity. We then consider the expectation value of the combination of the partial symmetry transformations as
\begin{align}
     \mathcal{T}^f_a\left(\theta,\phi\right):=\bra{\Psi_a}U_{\phi,\mathrm{A}}T_{\theta,\mathrm{A}} \ket{\Psi_a},
\end{align}
see Fig.~\ref{fig:rotation} for a schematic illustration.
The operator $T_{\theta,\mathrm{A}}$ is referred to as a partial rotation operator in this paper, while we note that it is called partial translation in previous work e.g., \cite{Qi2012momentumpolarization}.

Let us assume that $T_{\mathrm{A};\theta}$ induces the trivial symmetry fractionalization on the underlying topological order, meaning that the phase is trivial as a symmetry-enriched topological phase with translation symmetry.
We then find that this quantity with $(\theta,\phi) = (2\pi/n,\pi/n)$ solely gives the constant universal value independent of the system size; when $n$ is odd, we get
\begin{align} \begin{split}
    \mathcal{T}^f_a\left(\frac{2\pi}{n},\frac{\pi}{n}\right) & \propto e^{\textstyle\frac{2\pi i}{n}(\frac{\sigma_H}{8}-\frac{n^2+2}{24}c_-)} e^{\textstyle\frac{2\pi i}{n}h_a} \\&\qquad \times \sum_{b\in\mathcal{C}} d_b S_{ab}\theta_b^n e^{i\pi Q_b}.
\end{split} \end{align}
where $S_{ab}$ is the modular $S$ matrix of the bulk topological order, and $h_a$ is the spin carried by the vacuum in the topological sector of CFT labeled by the anyon $a$.
In particular, when $a=1$ and $n$ is odd,
\begin{align}
    \mathcal{T}^f_1\left(\frac{2\pi}{n},\frac{\pi}{n}\right)  & \propto e^{\frac{2\pi i}{n}(\frac{\sigma_H}{8}-(\frac{n^2+2}{24})c_-)}\sum_{a\in\mathcal{C}}e^{i\pi Q_a} d_a^2\theta_a^n,
    \label{eq:higherhall_partialrot}
    \end{align}
which is proportional to higher Hall conductivity $\zeta_n^H$ when $\gcd(n,\nfs)=1$, and defines the obstruction to $\U^f$-preserving gapped edge state beyond $c_-$ and $\sigma_H$.

One can do a simple consistency check of the formula \eqref{eq:higherhall_partialrot} 
by taking $n=1$. In that case, the unitary acting on the subsystem A is given by $U_{\pi,\mathrm{A}} T_{2\pi,\mathrm{A}}$. $U_{\pi}$ generates $\Z_2^f$ fermion parity symmetry, while $T_{2\pi}$ generates the $2\pi$ rotation of the cylinder. This cylinder is equipped with the anti-periodic boundary condition for $\Z_2^f$ symmetry, as this corresponds to the trivial sector $a=1$. Due to this boundary condition, $T_{2\pi}$ again amounts to the action of $\Z_2^f$ symmetry. As a whole, the combined operator $U_{\pi,\mathrm{A}} T_{2\pi,\mathrm{A}}$ gives an identity, hence one should have $\mathcal{T}^f_1\left(2\pi,\pi\right) =1$. This is satisfied with Eq.~\eqref{eq:higherhall_partialrot} thanks to $\zeta_1^H = e^{\frac{2\pi i}{8}(c_--\sigma_H)}$.

One can derive the above results of partial rotations using the cut-and-glue approach, which states that the reduced density matrix at the A subsystem is effectively described by the thermal density matrix of the edge CFT at high temperature near the bipartition~\cite{Qi2012entanglement, KitaevTEE}. 
This allows us to identify the expectation value of the partial symmetry operators in the bulk Hilbert space as that evaluated in the Hilbert space of edge CFT, and compute the quantity $\mathcal{T}^f_a\left(\frac{2\pi}{n},\frac{\pi}{n}\right)$ at the level of effective edge CFT. See Appendix~\ref{app:detailedCFT} for the detailed descriptions.

For completeness, let us also present the results when $n$ is even. We get
\begin{align}
\begin{split}
    \mathcal{T}^f_a\left(\frac{2\pi}{n},\frac{\pi}{n}\right) &\propto e^{\frac{2\pi i}{n}(\frac{\sigma_H}{8}-(\frac{n^2+2}{24})c_-)}\sum_{v'} e^{\frac{2\pi i}{n}(h_a + h_{v'})}
    \\
    &\quad \times \sum_{b\in \mathcal{C}}S_{ab} \theta_b^n e^{i\pi Q_b} S_{bv'},
    \end{split}
    \label{eq:Tfvortices}
\end{align}
where we sum over the labels of the $\Z_2^f$ vortex $v'$ carrying the smallest scaling dimension. $v'$ is regarded as an object of the minimal modular extension of the super-modular category $\mathcal{C}$.

While the above formula involves the sum over $\Z_2$ vortices $v'$, one gets a simpler formula valid for even $n$ when we do not include the partial U(1) transformation,
\begin{align}
\begin{split}
    \mathcal{T}^f_a\left(\frac{2\pi}{n},0\right) & \propto e^{-\frac{2\pi i}{24} (n+\frac{2}{n})c_-} e^{\frac{2\pi i}{n}h_a}  \sum_{b\in \mathcal{C}}S_{ab} \theta_b^n d_b.
    \end{split}
\end{align}
We note the viability of measuring the partial rotation. If the wavefunction is put on a universal quantum computer, we can evaluate the expectation value of the unitary operator through the Hadamard test as follows: Considering preparing the following state:

\begin{equation}
    H U^\mathrm{cont} \ket{+}_\mathrm{ancilla}\ket{\psi}
\end{equation}
where $H$ is the Hadarmard gate, $U^\mathrm{cont}$ is the controlled $U$ operation, $\ket{\pm}_\mathrm{ancilla} = (\ket{\uparrow}_\mathrm{ancilla} \pm \ket{\downarrow}_\mathrm{ancilla})/\sqrt{2}$ is an ancilla qubit, and $\ket{\psi}$ is the state of interest. Writing out the action of $U^\mathrm{cont}$ explicitly, the state we get is

\begin{equation}
    \frac{1}{\sqrt{2}}(\ket{+}_\mathrm{ancilla}\ket{\psi} + \ket{-}_\mathrm{ancilla} U\ket{\psi}).
\end{equation}

The $\sigma^z$ expectation value of the ancilla qubit is then given by $\mathrm{Re}(\braket{\psi|U|\psi})$, which is the real part of the expectation value. We can similarly measure the imaginary part by using $(\ket{\uparrow}_\mathrm{ancilla} -i \ket{\downarrow}_\mathrm{ancilla})/\sqrt{2}$ as the initial state on the ancilla register. By choosing $U$ to be the partial rotation operator, we can measure the expectation values straightforwardly on a quantum computer.

\subsubsection{Extraction of Hall conductivity via partial rotation in fermionic Abelian phases}
\label{subsubsec:extract hall fermion}

When the topological order is Abelian, one can extract both the electric and thermal Hall conductivity from the partial rotation and topological entanglement entropy.
For fermionic phases, the algebraic theory of anyons $\mathcal{C}$ is a super-modular category. In the case of Abelian phases, one can write the super-modular tensor category as $\mathcal{C} = \mathcal{C}_0\boxtimes\{1,\psi\}$ with a modular tensor category $\mathcal{C}_0$. Here $\boxtimes$ denotes the Deligne product, which physically implies that at the level of effective theory, fermionic Abelian phase is regarded as stacking of bosonic topological order $\mathcal{C}_0$ with fermionic invertible phase $\{1,\psi\}$.

With this in mind, let us illustrate the way of extracting $\sigma_H$ and $c_-$ of a fermionic Abelian phase. We assume we know the total quantum dimension $\mathcal{D}$ of $\mathcal{C}_0$ through e.g.~computing the topological entanglement entropy~\cite{KitaevTEE}.

When the phase is Abelian, the Frobenius-Schur exponent $\nfs$ must divide $2\mathcal{D}^2$. By using this fact and $\zeta_1^H = e^{\frac{2\pi i}{8}(c_--\sigma_H)}$, we can obtain $c_-$ and $\sigma_H$ from the following two quantities\footnote{We note that Eqs.~\eqref{eq:2D2+1}, \eqref{eq:2D2-1} do not fully determine the values of $c_-,\sigma_H$ as rational numbers, since the partial rotations depend on these quantities through the phases. However, given that the phases in Eqs.~\eqref{eq:2D2+1}, \eqref{eq:2D2-1} depend on $c_-,\sigma_H$ with the fractions of large denominators $\mathrm{lcm}(2\mathcal{D}^2-1,8)$, it is expected that these two equations fix $c_-,\sigma_H$ modulo the integers as large as the denominators $\mathrm{lcm}(2\mathcal{D}^2-1,8)$. If we want to get $c_-,\sigma_H$ with higher precisions, the partial rotations with rotation angles $2\pi/(2m\mathcal{D}^2\pm 1)$ with $m\in \mathbb{Z}$ to give the additional information to these quantities, and it is expected that we can get $c_-,\sigma_H$ with arbitrarily high precisions. It would be interesting to rigorously verify these exceptions.}
\begin{align}
    \begin{split}
    &\mathcal{T}^f_1\left(\frac{2\pi}{2\mathcal{D}^2+1},\frac{\pi}{2\mathcal{D}^2+1}\right) \\
    & \propto \exp\left[\tfrac{2\pi i}{2\mathcal{D}^2+1}\left(\tfrac{\sigma_H}{8}-\tfrac{(2\mathcal{D}^2+1)^2+2}{24}c_-\right) +\tfrac{2\pi i}{8}(c_-{-}\sigma_H)\right],
    \label{eq:2D2+1}
    \end{split}
\\
    \begin{split}
    &\mathcal{T}^f_1\left(\frac{2\pi}{2\mathcal{D}^2-1},\frac{\pi}{2\mathcal{D}^2-1}\right) \\
    &  \propto \exp\left[\tfrac{2\pi i}{2\mathcal{D}^2-1}\left(\tfrac{\sigma_H}{8}-\tfrac{(2\mathcal{D}^2-1)^2+2}{24}c_-\right) -\tfrac{2\pi i}{8}(c_-{-}\sigma_H)\right].
    \label{eq:2D2-1}
    \end{split}
\end{align}
That is, by obtaining the above two partial rotations and solving the linear equation with respect to $\sigma_H$, $c_-$ enables us to determine both electric and thermal Hall conductivity.

\subsection{Bosonic phase}

One can obtain similar results for (2+1)D bosonic topological phases with $\U$ symmetry. In that case, the quantity $\mathcal{T}^b_a\left(\theta,\phi\right):=\bra{\Psi_a}U_{\phi,\mathrm{A}}T_{\theta,\mathrm{A}} \ket{\Psi_a}$ gives a constant universal value for $(\theta,\phi) = (2\pi/n,2\pi m/n)$ with $n,m\in\Z$,
\begin{align} \begin{split}
    \mathcal{T}^b_a\left(\frac{2\pi}{n},\frac{2\pi m}{n}\right) & \propto e^{\frac{2\pi i}{n}(\frac{m^2}{2}\sigma_H-(\frac{n^2+2}{24})c_-)} e^{\frac{2\pi i}{n}h_a }
    \\
    &\qquad \times \sum_{b\in \mathcal{C}} d_bS_{ab} \theta_b^n e^{2i\pi m Q_b}.
\end{split} \end{align}
When $m=0$ and $a=1$, it reduces to the higher central charge of the bosonic topological phase~\cite{Kobayashi2024hcc}
\begin{align}
\begin{split}
    \mathcal{T}^b_a\left(\frac{2\pi}{n},0\right) & \propto e^{-\frac{2\pi i}{24} (n+\frac{2}{n})c_-} \zeta_n.
    \end{split}
\end{align}

\subsubsection{Extraction of Hall conductivity via partial rotation in bosonic Abelian phases}
\label{subsubsec:extract hall boson}
As we have done in the fermionic Abelian phases, one can also extract $\sigma_H$ and $c_-$ of the bosonic Abelian topological order from partial rotations and topological entanglement entropy.

First, we again compute topological entanglement entropy to obtain the total quantum dimension $\mathcal{D}$. 
Next, we obtain the chiral central charge $c_-$. By noticing that $2\mathcal{D}^2$ is divisible by $\nfs$ in Abelian phases, one can extract $c_-$ by the partial rotation
        \begin{align}
        \mathcal{T}^b_1\left(\frac{2\pi}{2\mathcal{D}^2},0\right) \propto e^{-2\pi i (\frac{2}{2\mathcal{D}^2}+2\mathcal{D}^2)\frac{c_-}{24}}~,   \end{align}
        from which we get $c_-$.

Then, we obtain the electric Hall conductivity $\sigma_H$ via partial rotation. To do this, pick an integer $k$ coprime with $2\mathcal{D}^2$, $\gcd(k,2\mathcal{D}^2) = 1$. The partial rotation with $(\theta,\phi)=(2\pi/k,2\pi/k)$ is then given by
    \begin{align}
\begin{split}
    \mathcal{T}^b_1\left(\frac{2\pi}{k},\frac{2\pi}{k}\right) &\propto  e^{\frac{2\pi i}{k}(\frac{\sigma_H}{2}-(\frac{k^2+2}{24})c_-)} \sum_{b\in \mathcal{C}} \theta_b^k M_{b,v} \\
    & = e^{\frac{2\pi i}{k}(\frac{\sigma_H}{2}-(\frac{k^2+2}{24})c_-)} \sum_{b\in \mathcal{C}} \theta_b^k M_{b,{\tilde v}^k}   \\
    & = e^{\frac{2\pi i}{k}(\frac{\sigma_H}{2}-(\frac{k^2+2}{24})c_-)} \sum_{b\in \mathcal{C}} \left(\frac{\theta_{b\times \tilde{v}}}{\theta_{\tilde{v}}}\right)^k \\
    &=e^{\frac{2\pi i}{k}(\frac{\sigma_H}{2}-(\frac{k^2+2}{24})c_-)} \frac{\zeta_n}{(\theta_{\tilde{v}})^k} ~,
    \end{split}
\end{align} 
where $M_{a,b}$ denotes the mutual braiding between anyons $a,b$, and
 $v$ is a specific anyon called a vison satisfying $M_{b,v} = e^{2i\pi Q_a}$, and the anyon $\tilde{v}$ satisfies $v=\tilde{v}^k$ which exists since $\gcd(k,2\mathcal{D}^2) = 1$. We then have
\begin{align}
    \frac{\mathcal{T}^b_1\left(\frac{2\pi}{k},\frac{2\pi}{k}\right)}{\mathcal{T}^b_1\left(\frac{2\pi}{k},0\right)} = \frac{e^{\frac{2\pi i}{k}\sigma_H}}{(\theta_{\tilde{v}})^k}. 
    \end{align}  
    Let us take an integer $l$ satisfying $kl=1$ mod $2\mathcal{D}^2$. Noticing that $\theta_v=e^{i\pi \sigma_H}$, we can extract $\sigma_H$ from  the  ratio of partial rotations
    \begin{align}
    \frac{\mathcal{T}^b_1\left(\frac{2\pi}{k},\frac{2\pi}{k}\right)}{\mathcal{T}^b_1\left(\frac{2\pi}{k},0\right)} =  e^{\pi i \sigma_H(\frac{1}{k}-l)}~, 
    \label{eq:bosonicsigmaH}
    \end{align}
from which we get $\sigma_H$. 

\section{Numerical results}
\label{sec:numerics}
We first briefly describe the procedure to compute the partial $\U$ rotation given a matrix product state (MPS). The technology we need here is very similar to the one used in Ref.~\onlinecite{Kobayashi2024hcc}, whereas the only difference is that we also need the particle number label for the $\U$ rotation. For the quantum Hall systems we consider 
here, we always work with an infinite cylindrical geometry, and Landau gauge orbitals around the cylinder. We conserve both momentum around the cylinder and particle number, allowing us to reach bond dimension $\chi=3200$ and cylinder circumference $L_y=40 \ell_B$ such that the partial rotation in all Abelian phases studied is well converged. We use the real-space entanglement spectrum (RSES) algorithm described in Ref.~\onlinecite{exactmps} to cut the cylinder in two halves in physical space, along which we perform the partial $\U$ rotation. The final expectation value we need can be evaluated at the virtual bond corresponding to the cut,
\begin{equation}
    \mathcal{T}_a(\theta,\phi)=\sum_\alpha \lambda_\alpha^2 e^{i k_y^\alpha L_y \theta} e^{i n^\alpha \phi}
\end{equation}
where $\lambda_\alpha$ is the Schmidt value, and $k_y^\alpha$ and $n^\alpha$ are the corresponding physical momentum and particle number. Here we resolve the ambiguity in the momentum label by matching the entanglement spectrum and the edge CFT spectrum, such that (a) the highest weight state has zero momentum and charge, and (b) that sectors with opposite charges have the same momentum labels.

Now we present $\mathcal{T}_1^f\left(\frac{2 \pi}{n}, \frac{\pi}{n}\right)$ of the fermionic $\U_m$ topological order, which have $2m$ Abelian anyons $\{[j]_{2m}\}$ for $0\le j\le 2m-1$  with $\theta_j=\exp(\pi i \, j^2/m)$, $Q_j = j/m$, following the $\Z_{2m}$ fusion rule. The anyon $[m]_{2m}$ corresponds to a transparent fermion $\psi$. 
In the expression of $\mathcal{T}_1^f\left(\frac{2 \pi}{n}, \frac{\pi}{n}\right)$ for even $n$ in Eq.~\eqref{eq:Tfvortices}, we need to sum over the $\Z_2$ vortices $v'$ with the smallest spin, which are $v'=([1]_{4m}),([4m-1]_{4m})$ with $h_{v'}=1/8m$ in $\U_{4m}$. Here, note that the $\Z_2$ vortices are described by the object of the minimal modular extension of $\U_m$, which is $\U_{4m}$.

\begin{figure}[htbp]
    \centering
    \includegraphics[width = 0.48\textwidth]{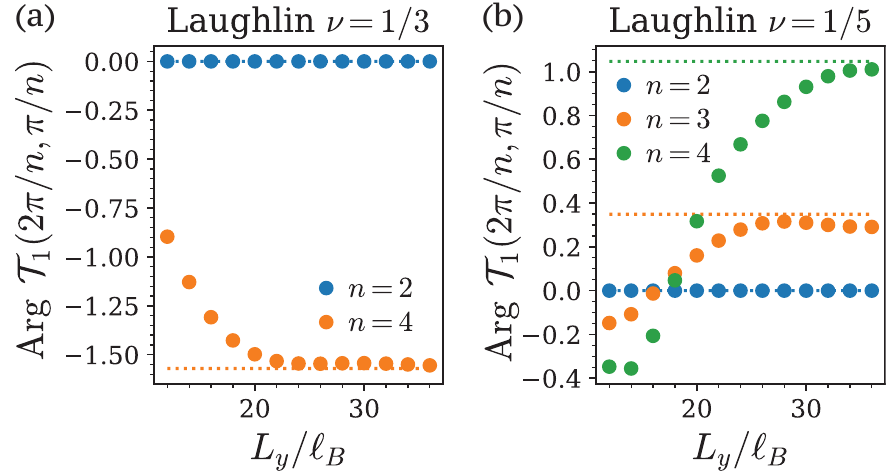}
    \caption{$\operatorname{Arg} \mathcal{T}_1(2\pi/n,\pi/n)$ of the $\nu = 1/3$ and $\nu = 1/5$ Laughlin states. The dotted lines are the CFT predictions in Table \ref{tab:fermionphases}.}
    \label{fig:Laughlin}    
\end{figure}

We show the expectation
values of $\mathcal{T}_1^f\left(\frac{2 \pi}{n}, \frac{\pi}{n}\right)$ for small $n$ in the $\nu=1/3,1/5$ Laughlin state in Table \ref{tab:fermionphases} as microscopic realization of the $\U_m$ topological order.
The model we study here is two-dimensional fermions in the lowest Landau level (LLL) at $\nu = 1/m$ filling with an interaction $V_{n < m} = 1$ plus a small perturbation $\delta V_{n = m} = 0.1$, where $V_n$ are the Haldane pseudopotentials \cite{Haldane_1994_pseudopotential,halperin_jain_cooper_2020}. We obtained the ground state MPS from infinite density matrix renormalization group (iDMRG) calculations~\cite{PhysRevB.91.045115}. As shown in Fig.~\ref{fig:Laughlin} (b), $\mathcal{T}_1^f\left(\frac{2 \pi}{n}, \frac{\pi}{n}\right)$ always converges to the expected phase in Table \ref{tab:fermionphases} at sufficiently large $L_y$.

\begin{figure}[htbp]
    \centering
    \includegraphics[width = 0.48\textwidth]{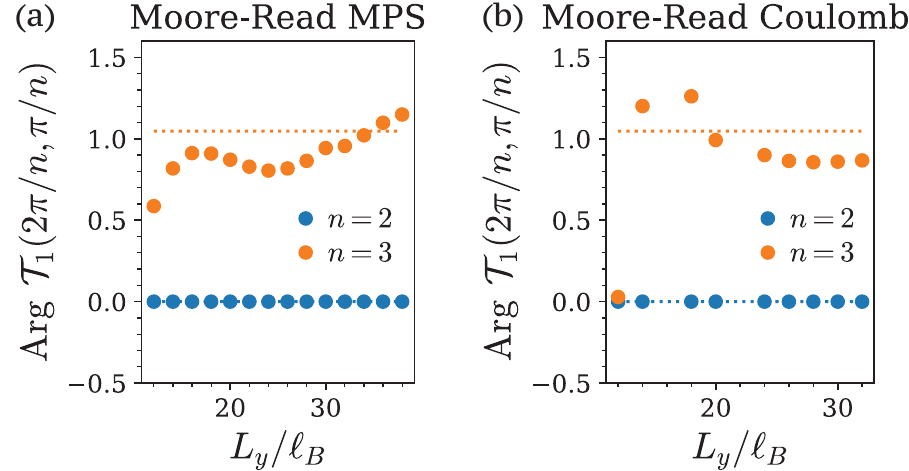}
    \caption{$\operatorname{Arg} \mathcal{T}_1(2\pi/n,\pi/n)$ of the $\nu= 1/2$ Moore-Read state. We obtain the ground state from (a) CFT construction and (b) iDMRG calculation respectively. The dotted lines are the CFT predictions given in Table \ref{tab:fermionphases}.}
    \label{fig:MR}    
\end{figure}

The non-Abelian phases, however, present more numerical challenges. Here we show $\mathcal{T}_1^f\left(\frac{2 \pi}{n}, \frac{\pi}{n}\right)$ of the Moore-Read state in Fig.~\ref{fig:MR}. The set of anyons in the Moore-Read state is obtained by condensing a fermion $\psi$ in the $\U_8\times \mathrm{Ising}$ topological order. The anyons are labeled by $([j]_8,a)$ with $a\in\{1,\sigma,\epsilon\}$, where the fermion is labeled by $\psi=([4]_8,\epsilon)$. The fractional U(1) charge is given by $Q_{([j]_8,a)}=j/4$.
See Table~\ref{tab:mooreread} for spins of the anyons in the Moore-Read state, and Table~\ref{tab:fermionphases} for the values of $\mathcal{T}_1^f\left(\frac{2 \pi}{n}, \frac{\pi}{n}\right)$ for small $n$. In the expression of $\mathcal{T}_1^f\left(\frac{2 \pi}{n}, \frac{\pi}{n}\right)$ for even $n$ in Eq.~\eqref{eq:Tfvortices}, we need to sum over the $\Z_2$ vortices $v'$ with the smallest spin, which are $v'=([1]_8,1),([7]_8,1),([0]_8,\sigma)$ with $h_m=1/16$ in $\U_8\times\mathrm{Ising}$.

We first construct the MPS for the Moore-Read model wavefunction from the corresponding $1+1$D chiral CFT, a tensor product of a chiral boson and a Majorana mode~\cite{exactmps}. Here we keep $17$ CFT Virasoro levels in the MPS construction. As shown in Fig.~\ref{fig:MR} (a), $\mathcal{T}_1^f\left(\frac{2 \pi}{n}, \frac{\pi}{n}\right)$ is converging, although oscillatory, to the expected phase in Table \ref{tab:fermionphases}. We cannot reach full convergence even at the largest system size we reach. We attribute the slow convergence to the longer correlation length of the Moore-Read state. To explore the Moore-Read phase, we also study a microscopic model of half-filled $N = 1$ Landau level of two-dimensional fermions with long-range Coulomb interaction using iDMRG. 
% We obtain the ground state MPS using iDMRG, which is known to have a $78\%$ overlap with the model wavefunction \cite{MRoverlap}.
% \roger{overlap doesn't make sense without context.  Why report overlap?}
As shown in Fig.~\ref{fig:MR} (b), $\mathcal{T}_1^f\left(\frac{2 \pi}{n}, \frac{\pi}{n}\right)$ is also converging to the expected values with this more generic state in the Moore-Read phase.

\begin{figure}[htbp]
    \centering
    \includegraphics[width = 0.48\textwidth]{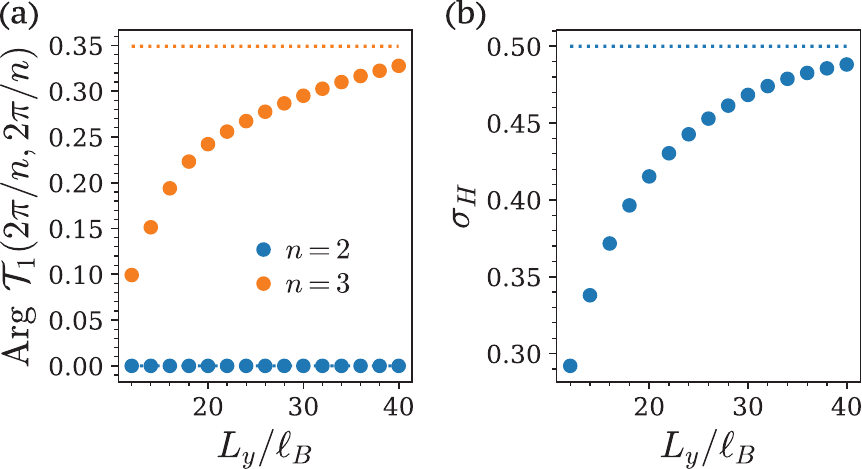}
    \caption{(a) $\operatorname{Arg} \mathcal{T}_1(2\pi/n,2\pi/n)$ of the $\nu= 1/2$ bosonic Laughlin state. 
    The dotted lines are the CFT predictions given in Table \ref{tab:bosonphases}. (b) Extracted electric Hall conductivity using Eq.~\eqref{eq:bosonicsigmaH}.}
    \label{fig:boson}    
\end{figure}

Finally, we show results of bosonic Abelian phases and explore the possibility to extract electric Hall conductance from partial rotation. Here we present $\mathcal{T}_1^b\left(\frac{2 \pi}{n}, \frac{2 \pi}{n}\right)$ of the bosonic $\nu =1/2$ Laughlin state, 
with a single non-trivial anyon being the semion. We follow Ref.~\onlinecite{Kobayashi2024hcc} to obtain the ground state MPS by solving the 
half-filled LLL of two-dimensional bosons with a contact interaction $V_0 = 1$ plus a small perturbation $\delta V_2 = 0.1$ with iDMRG. As shown in Fig.~\ref{fig:boson} (a), $\mathcal{T}_1^b\left(\frac{2 \pi}{n}, \frac{2 \pi}{n}\right)$ always converges to the expected phase as shown in Table \ref{tab:bosonphases} at sufficiently large $L_y$. Combining with $\mathcal{T}_1^b\left(\frac{2 \pi}{n}, 0\right)$ obtained in Ref.~\onlinecite{Kobayashi2024hcc}, we can extract the electric Hall conductance using Eq.~\eqref{eq:bosonicsigmaH}. As shown in Fig.~\ref{fig:boson} (b), the Hall conductance converges to the expected value $\sigma_H = 1/2$. The numerical success of Eq.~\eqref{eq:bosonicsigmaH} presents partial rotation as an appealing method to extract boson Hall conductance.

\begin{table}[]
\renewcommand*{\arraystretch}{1.5}
\centering
\begin{tabular}{c|c}
\hline \hline
        \   & $\mathcal{T}_1^f\left(\frac{2\pi}{n},\frac{\pi}{n}\right)$ \\ \hline
$\mathrm{U}(1)_3$     &  $1,1,0,-i$  \\ 
                         
{$\mathrm{U}(1)_5$}   & $1, 1, e^{\frac{13\pi i}{45}}(4 + 2 e^{2 \pi i/5} + 4e^{4 \pi i/5})\sim 4.20+1.53i, e^{\pi i/4} $  \\
MR & $1, 1 , e^{\pi i/3}, 0$ \\
\hline \hline
\end{tabular}
\caption{The phases of the partial rotation $\mathcal{T}^f_1(\frac{2\pi}{n},\frac{\pi}{n})$ for $n=1,2,3,4$ in the trivial sector of $\mathrm{U}(1)_3$, $\mathrm{U}(1)_5$ and Moore-Read. We write $0$ when the magnitude is vanishing.}
\label{tab:fermionphases}
\end{table}

\begin{table}[]
\renewcommand*{\arraystretch}{1.5}
\centering
\begin{tabular}{c|c|c}
\hline \hline
        \   & $\mathcal{T}_1^b\left(\frac{2\pi}{n},\frac{2\pi}{n}\right)$ &$\mathcal{T}_1^b\left(\frac{2\pi}{n},0\right)$ \\ \hline
$\mathrm{U}(1)_2$     & $1, 1, e^{\pi i/9}, 0 $  & $1, 0, e^{\frac{13\pi i}{9}}, e^{\frac{13\pi i}{8}}$  \\ 
\hline \hline
\end{tabular}
\caption{The phases of the partial rotation $\mathcal{T}^b_1(\frac{2\pi}{n},\frac{2\pi}{n}), \mathcal{T}_1^b\left(\frac{2\pi}{n},0\right)$ for $n=1,2,3,4$ in the trivial sector of the bosonic Laughlin state $\mathrm{U}(1)_2$. We write $0$ when the magnitude is vanishing.}
\label{tab:bosonphases}
\end{table}

\begin{table}[t]
	\centering
	\begin{tabular} {c ||c |c |c| c| c| c| c| c}
	\ & [0]$_8$ & [1]$_8$ & [2]$_8$ & [3]$_8$ & [4]$_8$ & [5]$_8$ & [6]$_8$ & [7]$_8$ \\
	\hline
	$1$ & 0 & \ & $\frac{1}{4}$ & \ & 0 & \ & $\frac{1}{4}$ & \ \\
	$\sigma$ &\ & $\frac{1}{8}$& \ & $\frac{5}{8}$ & \ & $\frac{5}{8}$ & \ & $\frac{1}{8}$ \\
	$\epsilon$ &$\frac{1}{2}$ & \ & $\frac{3}{4}$ & \ & $\frac{1}{2}$ & \ & $\frac{3}{4}$ &\ \\
      \end{tabular}
      \caption{Spins of anyons in the Moore-Read state.}
      \label{tab:mooreread}
\end{table}

\section{Complete set of obstructions to U(1)-preserving gapped edge in Abelian topological order}
\label{sec:complete}
In this section, we show that together with topological entanglement entropy, the invariants $\mathcal{T}^f_1,\mathcal{T}^b_1$ completely determine the gappability of the edge state in the case of Abelian topological order with $\U$ symmetry. This result holds for both fermionic and bosonic Abelian phases, and we provide an explicit algorithm for determining gappability in both cases.

\subsection{Bosonic Abelian phase}
Let us start with bosonic topological order with $\U$ symmetry. 
Together with topological entanglement entropy, one can see that the partial rotations $\mathcal{T}^b_1\left(\frac{2\pi}{n},0\right), \mathcal{T}^b_1\left(\frac{2\pi}{n},\frac{2\pi}{n}\right)$ completely determines if the bosonic Abelian topological order admits a $\U$-preserving gapped edge state or not. 

To see this, let us recall that higher central charges $\{\zeta_n\}$ with $n>1$ together with the electric and thermal Hall conductivity $\sigma_H, c_-$ can fully determine the U(1)-preserving edge gappability \cite{Kobayashi2022FQH}. 

It was already shown in Ref.~\onlinecite{Kobayashi2024hcc} that $\{\zeta_n\}$ can all be extracted from the partial rotation. 
In addition, both $\sigma_H$ and $c_-$ are shown to be extracted from partial rotation and topological entanglement entropy in Sec.~\ref{subsubsec:extract hall boson}. These span all the obstructions to $\U$-preserving gapped edge state mentioned above.

\subsection{Fermionic Abelian phase}
One can also completely determine if a fermionic Abelian topological order with U(1)$^f$ symmetry admits a symmetry-preserving gapped edge state or not, based on the partial rotation $\mathcal{T}^f_1\left(\frac{2\pi}{n},\frac{\pi}{n}\right), \mathcal{T}^f_1\left(\frac{2\pi}{n},0\right)$ and topological entanglement entropy.

For fermionic Abelian phases, one can write the super-modular tensor category as $\mathcal{C} = \mathcal{C}_0\boxtimes\{1,\psi\}$ with a modular tensor category $\mathcal{C}_0$.
One can then show that $c_-, \sigma_H$ and set of higher central charges $\{\zeta_n(\mathcal{C}_0)\}$ for $\gcd(n,\frac{\nfs(\mathcal{C}_0)}{\gcd(n,\nfs(\mathcal{C}_0) )})=1$ 
give the complete obstructions to the U(1)-preserving gapped edge state. See Appendix~\ref{app:abelianFQE} for a proof. 

In Sec.~\ref{subsubsec:extract hall fermion}, we have already seen that $\mathcal{D}$, $c_-,\sigma_H$ are extracted from partial rotation and topological entanglement entropy. Hence, the remaining task for determining gappability is to extract higher central charges $\{\zeta_n(\mathcal{C}_0)\}$ when $\sigma_H=c_-=0$.
Below, we provide an explicit algorithm for extracting $\{\zeta_n(\mathcal{C}_0)\}$ in Abelian fermionic phase, when $\sigma_H=c_-=0$ is understood and $\mathcal{D}$ is already known.

For convenience, let us write the prime factorization of $\mathcal{D}^2$ as
\begin{align}
    \mathcal{D}^2 = p_1^{k_1}\times p_2^{k_2}\times \dots \times p_M^{k_M}
\end{align}
with $p_1< p_2<\dots < p_M$ prime numbers, and $k_1,k_2,\dots,k_M$ positive integers. Then, the modular tensor category $\mathcal{C}_0$ also factorizes~\cite{kaidi2021higher} as
\begin{align}
    \mathcal{C}_0 = \mathcal{C}_{p_1} \boxtimes \mathcal{C}_{p_2} \boxtimes \dots \boxtimes  \mathcal{C}_{p_M}.
\end{align}
Each Abelian theory $\mathcal{C}_{p_j}$ contains the $p_j^{k_j}$ anyons including the trivial one. The Frobenius-Schur exponent also naturally admits the factorization as
\begin{align}
    \nfs(\mathcal{C}_0) = N_1N_2\dots N_{M},
\end{align}
where $N_j$ gives the Frobenius-Schur exponent of the theory $\mathcal{C}_{p_j}$. Each $N_j$ is expressed by a positive integer power of $p_j$, $N_j = p_j^{l_j}$, where $l_j$ satisfies
\begin{align}
    \begin{cases}
        1\le l_1 \le k_1+1 & \text{if $p_1=2$,} \\
        1\le l_j \le k_j & \text{otherwise}.
    \end{cases}
\end{align}
In particular, the above relations explain why $\nfs$ must divide $2\mathcal{D}^2$. 

Further, a vison $v$ encoding the U(1)$^f$ symmetry fractionalization also admits a factorization in $\mathcal{C} = \mathcal{C}_0\boxtimes\{1,\psi\}$ as
\begin{align}
    v = v_1\times v_2 \dots\times v_M \times v_\psi, 
\end{align}
where $v_j\in\mathcal{C}_{p_j}$, and $v_{\psi}$ is a vortex in the modular extension of $\{1,\psi\}$ satisfying $M_{v_{\psi},\psi}=-1$.

Now we are ready to describe all higher central charges in terms of the partial rotation. For this purpose, we note that having $\zeta_n(\mathcal{C}_0)=1$ for all $n$ such that $\gcd\bigl( n,\frac{\nfs(\mathcal{C}_0)}{\gcd(n,\nfs(\mathcal{C}_0) )} \bigr)=1$ is equivalent to having the higher central charge $\zeta_n(\mathcal{C}_{p_j}) = 1$ for all $j$ and $1\le n
\le N_j$ such that $\gcd(n,p_j) = 1$~\cite{kaidi2021higher}. 

One can cover all the above $\zeta_n(\mathcal{C}_{p_j})$ by computing either $\mathcal{T}^f_1(2\pi/n,\pi/n)$ or $\mathcal{T}^f_1(2\pi/n,0)$ with 
\begin{align}
    n = \frac{m\times 2\mathcal{D}^2}{p_j^{\tilde{k}_j}},
\end{align}
where $m$ scans all integers $1\le m\le p_j^{\tilde{k}_j}$ such that $\gcd(m,p_j)$ = 1. 
The integers $\tilde{k}_j$ are defined as
\begin{align}
    \begin{cases}
        \tilde{k}_1 = k_1+1 & \text{if $p_1 = 2$,} \\
        \tilde{k}_j=k_j & \text{otherwise}.
    \end{cases}
\end{align}
Here, depending on the value of $m$ we compute $\mathcal{T}^f_1(2\pi/n,\pi/n)$ when $n$ is odd, and compute $\mathcal{T}^f_1(2\pi/n,0)$ when $n$ is even.

Let us check that above partial rotations exhaust all desired higher central charges of $\mathcal{C}_{p_j}$, when $c_-=\sigma_H=0$. Since $2\mathcal{D}^2/p_j^{\tilde{k}_j}$ is coprime with $p_j$ and divisible by $N_i$ for all $i$ such that $i\neq j$, one can write the partial rotation with $n = m\times 2\mathcal{D}^2/p_j^{\tilde{k}_j}$ as
\begin{itemize}
\item when $n$ is odd and $c_-=\sigma_H=0$, 
\begin{align} \begin{split}
    & \mathcal{T}^f_1\left(\frac{2\pi}{n},\frac{\pi}{n}\right) \propto \sum_{a\in\mathcal{C}_0} e^{i\pi Q_a}\theta_a^n \\\qquad
    &= \prod_{1\le i\le M}\left(\sum_{a\in\mathcal{C}_{p_i}} e^{i\pi Q_a}\theta_a^n \right)  \propto\sum_{a\in\mathcal{C}_{p_j}} e^{i\pi Q_a}\theta_a^n \\\qquad
    & = \sum_{a\in \mathcal{C}_{p_j}} M_{a,v_j} \theta_a^n =\sum_{a\in \mathcal{C}_{p_j}} M_{a,\tilde v_j^n} \theta_a^n \\\qquad
    & =\sum_{a\in \mathcal{C}_{p_j}} \left(\frac{\theta_{a\times \tilde{v}_j}}{\theta_{\tilde{v}_j}}\right)^n\propto \frac{\zeta_n(\mathcal{C}_{p_j})}{\theta_{v_j}^l}.
\end{split} \end{align}
where $\tilde v_j\in \mathcal{C}_{p_j}$ satisfies $\tilde{v}_j^n = v_j$, and we take an integer $l$ satisfying $ln=1$ mod $p_j$. Since $\theta_{v_j}=1$ when $\sigma_H=0$, $\mathcal{T}^f_1\left(\frac{2\pi}{n},\frac{\pi}{n}\right)$ gives $\zeta_n(\mathcal{C}_{p_j})$ when $c_-,\sigma_H$ are vanishing. 

\item when $n$ is even and $c_-=0$, 
\begin{align}
\begin{split}
    \mathcal{T}^f_1\left(\frac{2\pi}{n},0\right) &\propto  \sum_{a\in\mathcal{C}_0} \theta_a^n \\
    &= \prod_{1\le i\le M}\left(\sum_{a\in\mathcal{C}_{p_i}} \theta_a^n \right)  \propto \sum_{a\in\mathcal{C}_{p_j}} \theta_a^n \\
    &\propto \zeta_n(\mathcal{C}_{p_j}).
    \end{split}
\end{align}

\end{itemize}
One can then see that $\zeta_n(\mathcal{C}_{p_j})$ covers all the desired higher central charges by scanning $m$, since $n$ mod $N_j$ covers all $1\le n \le N_j$ such that $\gcd(n,p_j)=1$.
Since we do not know $\nfs$ or $N_j$ in advance, we want to take the range of $m$ as $1\le m\le p_j^{\tilde{k}_j}$ so that $1\le m\le N_j$ is included.  This shows that the above $\mathcal{T}_1\left(2\pi/n\right)$ covers all gappability obstructions of $\mathcal{C}_{p_j}$.

Summarizing, we can obtain all gappability obstructions of the whole theory $\mathcal{C}$ beyond $c_-$ and $\sigma_H$, by computing the partial rotation 
($\mathcal{T}^f_1(2\pi/n, \pi/n)$ or $\mathcal{T}^f_1(2\pi/n, 0)$ depending on odd/even $n$) with
\begin{align}
    n = \frac{m\times 2\mathcal{D}^2}{p_j^{\tilde{k}_j}},
\end{align}
where $j$ scans all $1\le j\le M$, and $m$ scans all integers $1\le m\le p_j^{\tilde{k}_j}$ such that $\gcd(m,p_j)$ = 1.

\section{Lieb-Schultz-Mattis type constraint}
\label{sec:lsm}
In this section, we consider the setup where the topological order is not necessarily Abelian. We argue that the partial rotation can still be utilized to constrain the low-energy spectrum of the bulk-boundary system with $\U^f$ global symmetry, even for non-Abelian topological order.

Suppose that we observed $\{\TT^f_1(2\pi/p_j,\pi/p_j)\}$ has a non-trivial phase for a set of distinct prime numbers $\{p_j\}$.
One can see that this leaves us two possibilities: 1. the edge cannot be gapped while preserving $\U^f$ symmetry, or 2. the edge is gappable preserving $\U^f$ symmetry, where the bulk must have multiple ground states on a torus with the degeneracy non-trivially bounded from below. This statement is reminiscent of the Lieb-Schultz-Mattis type theorems~\cite{Lieb1961, Oshikawa, Hastings}, which constrains the low-energy spectrum for a given input of the symmetry action on the ground state. Concretely, one can show the following theorem: 

\textit{Theorem.} 
For a (2+1)D topological order with $\U^f$ symmetry, suppose that the partial rotation $\{\mathcal{T}^f_1(2\pi/p_j,\pi/p_j)\}$ has a non-trivial phase for a set of distinct prime numbers $\{p_j\}$.
Then, the topological order admits a $\U^f$ symmetry-preserving gapped edge state \textit{only if} the number of anyons $r$ satisfies $r\ge r_0$, where $r_0$ is the smallest integer satisfying $\prod_j p_j \le 2^{\frac{4}{3}r_0 + 8} 3^{\frac{4}{3}r_0}$. 

\textit{Proof.} 
Let us assume that the topological order admits a $\U^f$-preserving gapped edge, while the phases of  $\{\mathcal{T}_1^f(2\pi/p_j,2\pi/p_j)\}$ are non-trivial. $\mathcal{T}^f_1(2\pi/p_j,\pi/p_j)$ obstructs the $\U^f$-preserving gapped edge when $\gcd(\nfs(\mathcal{C}),p_j)=1$,  where $\mathcal{C}$ is the super-modular tensor category that describes the anyons of the topological order.
We hence must have $\gcd(\nfs(\mathcal{C}),p_j)>1$. This implies that $\nfs(\mathcal{C})$ must be divisible by $\prod_j p_j$, so we have $\prod_j p_j \le \nfs(\mathcal{C})$.
Meanwhile, let us consider a modular tensor category $\Breve{\mathcal{C}}$ obtained by gauging $\Z_2^f$ symmetry of the fermionic topological phase $\mathcal{C}$. Mathematically, $\Breve{\mathcal{C}}$ is the minimal modular extension of the super-modular category $\mathcal{C}$. For a modular category, it is known that
$\nfs(\Breve{\mathcal{C}}) \le 2^{\frac{2\Breve{r}}{3}+8}3^{\frac{2}{3}\Breve{r}}$, where $\Breve{r}$ is the number of anyons in $\Breve{\mathcal{C}}$~\cite{Bruillard2015rankfiniteness}. In the presence of $\U^f$ symmetry, there is an Abelian vison $v$ that carries $\Z_2^f$ vortex. In that case, one can write $\Breve{r} = 2r$, where $r$ is the number of anyons in $\mathcal{C}$. Using $\nfs(\mathcal{C}) \le\nfs(\Breve{\mathcal{C}})$, we have
\begin{align}
    \prod_j p_j \le \nfs(\mathcal{C})  \le\nfs(\Breve{\mathcal{C}})\le 2^{\frac{2\Breve{r}}{3}+8}3^{\frac{2}{3}\Breve{r}} = 2^{\frac{4}{3}r + 8}3^{\frac{4}{3}r}
\end{align}
which finishes the proof. \hfill$\square$

\section{Obstructions to gapped edge preserving Lie group symmetry}
\label{sec:lie}
Finally, let us provide a generalization of higher Hall conductivity to the case where the global symmetry is a generic connected Lie group. 
The idea for the generalization is straightforward; it is done by restricting the symmetry group to $\mathrm{U(1)}^f\subset G_f$ and defining higher Hall conductivity for each inequivalent choice of the subgroup $\U^f\subset G_f$.
This generalized version of higher Hall conductivity again gives a set of obstructions to gapped edge state preserving the Lie group symmetry, and can be extracted from partial rotation.

\subsection{Review: symmetry fractionalization of the Lie group}
To start with, let us outline how the Lie group symmetry generally acts on the topological order. Since the higher Hall conductivity and its generalizations are defined for fermionic systems, we consider the (2+1)D fermionic topological order with $G_f$ global symmetry. At the level of effective field theory, the anyons are thought to be simple objects of a super-modular category $\mathcal{C}$. Also, the anyons and $\Z_2^f$ vortices are the objects of a minimal modular extension $\Breve{\mathcal{C}}$ of the category $\mathcal{C}$. Physically, $\Breve{\mathcal{C}}$ is the bosonic theory obtained by gauging $\Z_2^f$ symmetry of the fermionic theory $\mathcal{C}$.

The symmetry group $G_f$ has the structure of the group extension
\begin{align}
    \Z_2^f\to G_f\to G_b,
\end{align}
with $\Z_2^f$ the fermion parity symmetry.
The bosonic symmetry group $G_b$ is assumed to be a connected Lie group, and the above group extension is specified by the element $[\omega_2]\in \mathcal{H}^2(G_b,\Z_2)$.

In general, it is convenient to specify $\omega_2$ by a choice of a group homomorphism $\pi_f: K\to \Z_2^f$ with $K:= \pi_1(G_b)$. That is, there is a group extension to the universal cover $\widetilde{G}_b$ of $G_b$ given by $[w_2]\in \mathcal{H}^2(G_b,K)$,
\begin{align}
    K\to \widetilde{G}_b\to G_b,
\end{align}
and then $\omega_2$ can be specified by
\begin{align}
    \omega_2({\bf g}, {\bf h}) = \pi_f(w_2({\bf g}, {\bf h})),
\end{align}
with ${\bf g}, {\bf h}\in G_b$.

The action of the global symmetry on topological order is then determined by the symmetry fractionalization pattern~\cite{barkeshli2019}. 

The connected Lie group symmetry cannot permute the labels of the anyons.
In that case, the equivalence class of the symmetry fractionalization is specified by the second cohomology $[\mathfrak{t}]\in \mathcal{H}^2(G_b,\mathcal{A})$~\cite{Bulmash2022fermfrac}, which encodes the assignment of fractional charge on anyons.
Here, $\mathcal{A}$ is the Abelian group for the fusion algebra of Abelian objects in $\Breve{\mathcal{C}}$, i.e., the group of Abelian anyons and Abelian $\Z_2^f$ vortices. 
This cohomology class $[\mathfrak{t}]$ is again characterized by the choice of a group homomorphism $\pi_{\mathcal{A}}: K\to\mathcal{A}$ via
\begin{align}
    \mathfrak{t}({\bf g}, {\bf h}) = \pi_{\mathcal{A}}(w_2({\bf g}, {\bf h})),
\end{align}
which is subject to a constraint~\cite{Bulmash2022fermfrac}
\begin{align}
    M_{\psi,\mathfrak{t}({\bf g}, {\bf h})} = (-1)^{\omega_2({\bf g}, {\bf h})},
\end{align}
where $M_{a,b}$ is the mutual braiding between $a,b$, and $\psi$ is a transparent fermion. This condition enforces that the symmetry fractionalization class $[\mathfrak{t}]$ is compatible with the fermionic symmetry group $G_f$, where the $G_b$ symmetry acts projectively on the transparent fermion $\psi$.

\subsection{Review: anomaly and Hall conductivity of topological order with Lie group symmetry}

Now we have set the symmetry action on the theory, then we can talk about the 't Hooft anomaly and generalized version of Hall conductivity of the topological order.
Suppose that the Abelian group $K$ is written as $K = \prod_{j} \Z_{n_j}$, and let ${\bf k}_j\in K$ be the generator of each $\Z_{n_j}$ group. 
In that case, the symmetry fractionalization is fully characterized by a set of anyons $\{\pi_{\mathcal{A}}({\bf k}_j)\}$; it fixes a map $\pi_{\mathcal{A}}$, so characterizes a symmetry fractionalization class $[\mathfrak{t}]$ accordingly. Physically, the anyon $\{\pi_{\mathcal{A}}({\bf k}_j)\}$ corresponds to the vortex whose vorticity is given by ${\bf k}_j\in \pi_1(G_b)$. When an anyon moves around the vortex, it is fractionally charged by the phase $M_{a,\pi_{\mathcal{A}}({\bf k}_j)}$ due to the Aharonov-Bohm effect.

Then, the Hall conductivity and anomalies of the theory can be characterized through the spins and braiding of the vortices $\{\pi_{\mathcal{A}}({\bf k}_j)\}$~\cite{Benini2019_2group, Cheng2023gauging}. 
Concretely, the (2+1)D topological order has the 't Hooft anomaly of $G_f$ symmetry characterized by the theta angle in (3+1)D (which may either be continuous or discrete),
\begin{align}
    \begin{split}
    & \sum_j 2\pi  h(\pi_{\mathcal{A}}({\bf k}_j)) \int_{W^4} \mathcal{P}(\pi_{\mathcal{A}}(w_2^j))
    \\& + \sum_{j<k} 2\pi  m(\pi_{\mathcal{A}}({\bf k}_j),\pi_{\mathcal{A}}({\bf k}_k))\int_{W^4} \pi_{\mathcal{A}}(w_2^j)\cup \pi_{\mathcal{A}}(w_2^k)~,
    \label{eq:anomalyliegroup}
    \end{split}
\end{align}
where $e^{2\pi i h(a)}=\theta_a$, $e^{2\pi i m(a,b)} = M_{a,b}$, and $\mathcal{P}$ is the Pontryagin square. 
$[w_2^j]\in \mathcal{H}^2(G_b,\Z_{n_j})$ labels the second Stiefel-Whitney class $[w_2]\in \mathcal{H}^2(G_b,K)$. 
Since the global symmetry $G_b$ is extended by fermion parity symmetry, the background gauge field of $G_b$ follows the shifted Dirac quantization condition~\cite{thorngren2019anomalies}
\begin{align}
    \sum_j \int_C \pi_f(g_b^\ast w_2^j) = \int_{C}w_2(TW^4)
    \quad \bmod 2,
    \label{eq:dirac}
\end{align}
for any 2-cycle $C$. Here $g_b: W^4\to BG_b$ is the background $G_b$ gauge field, and $g_b^*w_2^j\in H^2(W^4,\Z_{n_j})$ denotes the pull back of $w_2^j$.
$w_2(TW^4)$ is the second Stiefel-Whitney class of the tangent bundle of the spacetime $W^4$ on which the response action is supported. 

When the 't Hooft anomaly \eqref{eq:anomalyliegroup} defines a non-trivial SPT phase with $G_f$ symmetry, this means that our (2+1)D topological order has a non-trivial 't Hooft anomaly. We do not try to define the Hall conductivity in this case. Also, non-trivial 't Hooft anomaly automatically implies that the edge state of the (2+1)D topological order cannot be $G_f$ symmetric~\cite{Thorngren2021end}. Hence, we cannot talk about the generalization of higher Hall conductivity as an obstruction to $G_f$ symmetric gapped boundary. Since we are interested in the $G_f$ symmetric edge state, let us consider the case where 't Hooft anomaly of $G_f$ symmetry is absent. 

Then, let us assume that the 't Hooft anomaly in Eq.~\eqref{eq:anomalyliegroup} gives a trivial SPT phase. In that case, the action~\eqref{eq:anomalyliegroup} corresponds to a continuous theta angle rather than a discrete theta angle.
In that case, the (3+1)D action~\eqref{eq:anomalyliegroup} is identified as having the Chern-Simons type response action in (2+1)D, and describes the generalized version of Hall conductivity for $G_f$ symmetry. By fixing a suitable normalization, we can define the Hall conductivity through the level of the Chern-Simons action that corresponds to the theta angle~\eqref{eq:anomalyliegroup}.
As we will see below, we can also define a generalization of higher Hall conductivity, which is again associated with partial rotation of the wave function on a cylinder.

\subsection{Higher Hall conductivity and partial rotation of the topological order}

As mentioned above, let us assume that the 't Hooft anomaly of the (2+1)D topological order is trivial.
In this setup, one can again compute the partial rotation on a cylinder using the same logic as the case $G_f=\U^f$. We write the partial rotation associated with internal $G_f$ transformation as
\begin{align}
     \mathcal{T}_a\left(\theta;{\bf g}_f\right):=\bra{\Psi_a}U_{{\bf g}_f,\mathrm{A}}T_{\theta,\mathrm{A}} \ket{\Psi_a}
\end{align}
where ${\bf g}_f\in G_f$.

Suppose that $n$-th power of ${\bf g}_f$ becomes the fermion parity, ${\bf g}_f^n = (-1)^F$.
When $n$ is odd, one can then compute the partial rotation for the ${\bf g}_f$ as
\begin{align} \begin{split}
    \mathcal{T}_1\left(\frac{2\pi}{n}; {\bf g}_f\right) & \propto e^{-\frac{2\pi i}{24}(n+\frac{2}{n})c_-}e^{\frac{2\pi i}{8n}\sigma_H} \\
    &\qquad \times \sum_{b\in\mathcal{C}_0}   M_{b,\pi_\mathcal{A}({\bf k})} d_b^2 \theta_b^n~,
    \end{split}
    \label{eq:partialrotation_liegroup}
\end{align}
where ${\bf k}\in K$ is defined as follows. First let us define $\mathbf{g}\in G_b$ as the projection of  ${\bf g}_f$ onto the bosonic symmetry group $G_b$.
When $G_b$ is extended to $\tilde{G}_b$, ${\bf g}^n$ generally defines an element of $K$, and ${\bf k}\in K$ is defined as ${\bf g}^n = {\bf k}\in K$. 

$\sigma_H$ is the electric Hall conductivity defined by restricting the symmetry group to $\U^f$ symmetry that contains ${\bf g}_f$, and then consider the electric Hall conductivity of the $\U^f$ symmetric topological order.

When $\gcd(n,\nfs)=1$ is satisfied, this partial rotation defines an obstruction to gapped boundary preserving $G_f$ symmetry. 
One can define the above higher Hall conductivity for each choice of a loop $U(1)\subset G_b$ which corresponds to an element $\mathbf{k}\in K=\pi_1(G_b)$ satisfying $\pi_f(\mathbf{k}) = 1\in \Z_2^f$. 
Homotopically different choices of such loops (i.e., choices of $\mathbf{k}\in K$) generally give independent obstructions to a symmetric gapped edge state.

The computation of the partial rotation can be performed in the same fashion as the $\U^f$ case in terms of edge CFT (Appendix~\ref{app:detailedCFT}), since we only use the specific $\U^f$ subgroup of the whole Lie group symmetry.
We note that the CFT computation assumes the entanglement spectrum of the edge state preserving $G_f$ symmetry, which makes sense only if the bulk topological order does not have an 't Hooft anomaly~\cite{Thorngren2021end}. 

\subsection{Example: \texorpdfstring{$G_f = \mathrm{SU}(2)^f$}{Gf = SU(2)f}}

Let us consider the case where $G_f = \mathrm{SU}(2)^f$, $G_b = \mathrm{SO}(3)$. In that case, we have $K=\Z_2$, and the vortex $\pi_{\mathcal{A}}({\bf k})$ for SO(3) symmetry follows the $\Z_2$ fusion rule. The spin of $\pi_{\mathcal{A}}({\bf k})$ is given by $k/4$ with $k\in\Z_4$, and the bulk response action Eq.~\eqref{eq:anomalyliegroup} is given by
\begin{align}
    \frac{2\pi k}{4}\int_{W^4}\mathcal{P}(w_2(\mathrm{SO}(3))).
    \label{eq:SO3response}
\end{align}
This response action of $\mathrm{SU}(2)^f$ symmetry actually corresponds to a trivial SPT, when the Dirac quantization condition \eqref{eq:dirac} is understood; $w_2(\mathrm{SO}(3))= w_2(TW^4)$.
In that case we have $\int \mathcal{P}(w_2(TW^4))=\sigma(W^4)$ mod 4, with $\sigma(W^4)$ the signature of a 4-manifold $W^4$. Hence, the above response action~\eqref{eq:SO3response} is identified as the continuous gravitational theta angle, and topological order does not carry 't Hooft anomaly of SU(2)$^f$ symmetry.

Then, the partial rotation is given by Eq.~\eqref{eq:partialrotation_liegroup}. We have
$k/4 = \sigma_H/8$ mod 1, since the vison $\pi_{\mathcal{A}}(\mathbf{k})$ characterizes the fractionalization of $\U^f\in\mathrm{SU}(2)^f$ symmetry, and its spin is proportional to electric Hall conductivity as $h_{\pi_{\mathcal{A}}(\mathbf{k})} = \sigma_H/8$ mod 1~\cite{manjunath2020FQH}.
Eq.~\eqref{eq:partialrotation_liegroup} defines an obstruction to $\mathrm{SU}(2)^f$ preserving gapped edge state beyond $c_-$.

\section{Discussions}
\label{sec:discussion}
In this paper, we characterized the higher Hall conductivity $\zeta_n^H$ on a single wave function of fermionic topological order with $\U^f$ symmetry. $\zeta_n^H$ is extracted by evaluating the expectation value of the partial rotation followed by the partial $\U$ transformation. In the case of Abelian topological order, the partial rotation also enables us to extract $c_-$ and $\sigma_H$ from 
a single wave function, in both the bosonic and fermionic cases. Moreover, using partial rotations one can completely determine if the given Abelian topological order admits a symmetry-preserving gapped edge state. The higher Hall conductivity from partial rotation can generally be used to constrain the low-energy spectrum of the bulk-boundary system in possibly non-Abelian topological order with $\U^f$ symmetry, and its natural generalizations to the case with Lie group symmetry are 
also presented.

Let us conclude our paper with several open questions to be addressed in future work.
First of all, in a bosonic non-Abelian topological order, it is known that $\zeta_n$ only gives necessary conditions for gapped boundary and in general not sufficient. For instance, 20 copies of Fibonacci topological order (stacked with $\overline{E}_8$ states) does not admit a gapped edge state, even though it carries $c_-=0$ as well as $\zeta_n=1$ for all $n$~\cite{Neupert2016nogo, kaidi2021higher}. This calls for an easily computable quantity that can detect ingappability of edge state beyond higher central charges.

Second, while we only discussed the fermionic phases with $\U^f$ symmetry, it was recently found that the non-Abelian fermionic topological order without any additional global symmetry has a set of invariants which gives obstructions to a gapped edge state~\cite{you2023gapped}. Similar to higher central charges of Abelian topological order, these invariants are also computable in terms of the modular $S, T$ matrices of
a given topological order. A microscopic characterization of these invariants on a single many-body wave function is a fruitful future direction.

Also, recent development of non-invertible symmetries uncovered the presence of various invariants associated with the responses of non-invertible symmetries~\cite{thorngren2019fusioncategorysymmetryi, Zhang:2023wlu, Cordova:2023bja}. In particular, the electric Hall conductance can be defined for certain class of continuous non-invertible symmetries, which again forbids gapless edge modes preserving the symmetry~\cite{Hsin2024coset}. It would be interesting to extract such invariants from a microscopic wave function, as well as find higher versions of Hall conductivity for non-invertible symmetries.

Finally, the partial operations also allow 
us to characterize the invariants of crystalline topological phases with point group symmetry~\cite{Shiozaki2017point,Zhang2023partial, kobayashi2024RT}. It would be interesting to develop a theory for the crystalline invariants of topological order, which is relevant to various invariants characterizing the crystalline spin liquids.

\begin{acknowledgments}
R.K.\ is supported by JQI postdoctoral fellowship at the University of Maryland, and by National Science Foundation QLCI grant OMA-2120757.
T.W.\ is supported by the U.S.\ Department of Energy, Office of Science, Office of Basic Energy Sciences, Materials Sciences and Engineering Division under Contract No.\ DE-AC02-05-CH11231 (Theory of Materials program KC2301).
T.W. is also supported by the Heising-Simons Foundation, the Simons Foundation, and NSF grant No. PHY-2309135 to the Kavli Institute for Theoretical Physics (KITP). 
This research used the Lawrencium computational cluster provided by the Lawrence Berkeley National Laboratory (Supported by the U.S. Department of Energy, Office of Basic Energy Sciences under Contract No.\ DE-AC02-05-CH11231).
This research is funded in part by the
Gordon and Betty Moore Foundation’s EPiQS Initiative,
Grant GBMF8683 to T.S.
R.M.\ is supported by National Science Foundation grant DMR-1848336.
S.R.\ is supported by a Simons Investigator Grant from
the Simons Foundation (Award No.~566116).
This work is supported by
the Gordon and Betty Moore Foundation through Grant
GBMF8685 toward the Princeton theory program. 

\end{acknowledgments}

% \bibliography{bibliography.bib}

% \appendix

\bibliography{bibliography}

%merlin.mbs apsrev4-1.bst 2010-07-25 4.21a (PWD, AO, DPC) hacked
%Control: key (0)
%Control: author (8) initials jnrlst
%Control: editor formatted (1) identically to author
%Control: production of article title (-1) disabled
%Control: page (0) single
%Control: year (1) truncated
%Control: production of eprint (0) enabled
\begin{thebibliography}{62}%
\makeatletter
\providecommand \@ifxundefined [1]{%
 \@ifx{#1\undefined}
}%
\providecommand \@ifnum [1]{%
 \ifnum #1\expandafter \@firstoftwo
 \else \expandafter \@secondoftwo
 \fi
}%
\providecommand \@ifx [1]{%
 \ifx #1\expandafter \@firstoftwo
 \else \expandafter \@secondoftwo
 \fi
}%
\providecommand \natexlab [1]{#1}%
\providecommand \enquote  [1]{``#1''}%
\providecommand \bibnamefont  [1]{#1}%
\providecommand \bibfnamefont [1]{#1}%
\providecommand \citenamefont [1]{#1}%
\providecommand \href@noop [0]{\@secondoftwo}%
\providecommand \href [0]{\begingroup \@sanitize@url \@href}%
\providecommand \@href[1]{\@@startlink{#1}\@@href}%
\providecommand \@@href[1]{\endgroup#1\@@endlink}%
\providecommand \@sanitize@url [0]{\catcode `\\12\catcode `\$12\catcode
  `\&12\catcode `\#12\catcode `\^12\catcode `\_12\catcode `\%12\relax}%
\providecommand \@@startlink[1]{}%
\providecommand \@@endlink[0]{}%
\providecommand \url  [0]{\begingroup\@sanitize@url \@url }%
\providecommand \@url [1]{\endgroup\@href {#1}{\urlprefix }}%
\providecommand \urlprefix  [0]{URL }%
\providecommand \Eprint [0]{\href }%
\providecommand \doibase [0]{http://dx.doi.org/}%
\providecommand \selectlanguage [0]{\@gobble}%
\providecommand \bibinfo  [0]{\@secondoftwo}%
\providecommand \bibfield  [0]{\@secondoftwo}%
\providecommand \translation [1]{[#1]}%
\providecommand \BibitemOpen [0]{}%
\providecommand \bibitemStop [0]{}%
\providecommand \bibitemNoStop [0]{.\EOS\space}%
\providecommand \EOS [0]{\spacefactor3000\relax}%
\providecommand \BibitemShut  [1]{\csname bibitem#1\endcsname}%
\let\auto@bib@innerbib\@empty
%</preamble>
\bibitem [{\citenamefont {Halperin}(1982)}]{PhysRevB.25.2185}%
  \BibitemOpen
  \bibfield  {author} {\bibinfo {author} {\bibfnamefont {B.~I.}\ \bibnamefont
  {Halperin}},\ }\href {\doibase 10.1103/PhysRevB.25.2185} {\bibfield
  {journal} {\bibinfo  {journal} {Phys. Rev. B}\ }\textbf {\bibinfo {volume}
  {25}},\ \bibinfo {pages} {2185} (\bibinfo {year} {1982})}\BibitemShut
  {NoStop}%
\bibitem [{\citenamefont {Wen}(1990)}]{PhysRevLett.64.2206}%
  \BibitemOpen
  \bibfield  {author} {\bibinfo {author} {\bibfnamefont {X.~G.}\ \bibnamefont
  {Wen}},\ }\href {\doibase 10.1103/PhysRevLett.64.2206} {\bibfield  {journal}
  {\bibinfo  {journal} {Phys. Rev. Lett.}\ }\textbf {\bibinfo {volume} {64}},\
  \bibinfo {pages} {2206} (\bibinfo {year} {1990})}\BibitemShut {NoStop}%
\bibitem [{\citenamefont {Stone}(1991)}]{Stone:1990iw}%
  \BibitemOpen
  \bibfield  {author} {\bibinfo {author} {\bibfnamefont {M.}~\bibnamefont
  {Stone}},\ }\href {\doibase 10.1016/0003-4916(91)90177-A} {\bibfield
  {journal} {\bibinfo  {journal} {Annals Phys.}\ }\textbf {\bibinfo {volume}
  {207}},\ \bibinfo {pages} {38} (\bibinfo {year} {1991})}\BibitemShut
  {NoStop}%
\bibitem [{\citenamefont {Frohlich}\ and\ \citenamefont
  {Kerler}(1991)}]{Frohlich:1990xz}%
  \BibitemOpen
  \bibfield  {author} {\bibinfo {author} {\bibfnamefont {J.}~\bibnamefont
  {Frohlich}}\ and\ \bibinfo {author} {\bibfnamefont {T.}~\bibnamefont
  {Kerler}},\ }\href {\doibase 10.1016/0550-3213(91)90360-A} {\bibfield
  {journal} {\bibinfo  {journal} {Nucl. Phys. B}\ }\textbf {\bibinfo {volume}
  {354}},\ \bibinfo {pages} {369} (\bibinfo {year} {1991})}\BibitemShut
  {NoStop}%
\bibitem [{\citenamefont {Hatsugai}(1993)}]{hatsugai1993}%
  \BibitemOpen
  \bibfield  {author} {\bibinfo {author} {\bibfnamefont {Y.}~\bibnamefont
  {Hatsugai}},\ }\href {\doibase 10.1103/PhysRevLett.71.3697} {\bibfield
  {journal} {\bibinfo  {journal} {Phys. Rev. Lett.}\ }\textbf {\bibinfo
  {volume} {71}},\ \bibinfo {pages} {3697} (\bibinfo {year}
  {1993})}\BibitemShut {NoStop}%
\bibitem [{\citenamefont {Kane}\ and\ \citenamefont
  {Fisher}(1997)}]{KaneFisher}%
  \BibitemOpen
  \bibfield  {author} {\bibinfo {author} {\bibfnamefont {C.~L.}\ \bibnamefont
  {Kane}}\ and\ \bibinfo {author} {\bibfnamefont {M.~P.~A.}\ \bibnamefont
  {Fisher}},\ }\href {\doibase 10.1103/PhysRevB.55.15832} {\bibfield  {journal}
  {\bibinfo  {journal} {Phys. Rev. B}\ }\textbf {\bibinfo {volume} {55}},\
  \bibinfo {pages} {15832} (\bibinfo {year} {1997})}\BibitemShut {NoStop}%
\bibitem [{\citenamefont {Kitaev}(2006)}]{Kitaevanyons}%
  \BibitemOpen
  \bibfield  {author} {\bibinfo {author} {\bibfnamefont {A.}~\bibnamefont
  {Kitaev}},\ }\href {\doibase 10.1016/j.aop.2005.10.005} {\bibfield  {journal}
  {\bibinfo  {journal} {Annals of Physics}\ }\textbf {\bibinfo {volume}
  {321}},\ \bibinfo {pages} {2} (\bibinfo {year} {2006})},\ \Eprint
  {http://arxiv.org/abs/cond-mat/0506438} {arXiv:cond-mat/0506438} \BibitemShut
  {NoStop}%
\bibitem [{\citenamefont {Mitchell}\ \emph {et~al.}(2018)\citenamefont
  {Mitchell}, \citenamefont {Nash}, \citenamefont {Hexner}, \citenamefont
  {Turner},\ and\ \citenamefont {Irvine}}]{Mitchell2018amorphus}%
  \BibitemOpen
  \bibfield  {author} {\bibinfo {author} {\bibfnamefont {N.~P.}\ \bibnamefont
  {Mitchell}}, \bibinfo {author} {\bibfnamefont {L.~M.}\ \bibnamefont {Nash}},
  \bibinfo {author} {\bibfnamefont {D.}~\bibnamefont {Hexner}}, \bibinfo
  {author} {\bibfnamefont {A.~M.}\ \bibnamefont {Turner}}, \ and\ \bibinfo
  {author} {\bibfnamefont {W.~T.~M.}\ \bibnamefont {Irvine}},\ }\href {\doibase
  10.1038/s41567-017-0024-5} {\bibfield  {journal} {\bibinfo  {journal} {Nature
  Physics}\ }\textbf {\bibinfo {volume} {14}},\ \bibinfo {pages} {380}
  (\bibinfo {year} {2018})}\BibitemShut {NoStop}%
\bibitem [{\citenamefont {Tu}\ \emph {et~al.}(2013)\citenamefont {Tu},
  \citenamefont {Zhang},\ and\ \citenamefont
  {Qi}}]{Qi2012momentumpolarization}%
  \BibitemOpen
  \bibfield  {author} {\bibinfo {author} {\bibfnamefont {H.-H.}\ \bibnamefont
  {Tu}}, \bibinfo {author} {\bibfnamefont {Y.}~\bibnamefont {Zhang}}, \ and\
  \bibinfo {author} {\bibfnamefont {X.-L.}\ \bibnamefont {Qi}},\ }\href
  {\doibase 10.1103/physrevb.88.195412} {\bibfield  {journal} {\bibinfo
  {journal} {Physical Review B}\ }\textbf {\bibinfo {volume} {88}} (\bibinfo
  {year} {2013}),\ 10.1103/physrevb.88.195412},\ \Eprint
  {http://arxiv.org/abs/arXiv:1212.6951} {arXiv:1212.6951} \BibitemShut
  {NoStop}%
\bibitem [{\citenamefont {Zaletel}\ \emph {et~al.}(2013)\citenamefont
  {Zaletel}, \citenamefont {Mong},\ and\ \citenamefont {Pollmann}}]{FQHEDMRG}%
  \BibitemOpen
  \bibfield  {author} {\bibinfo {author} {\bibfnamefont {M.~P.}\ \bibnamefont
  {Zaletel}}, \bibinfo {author} {\bibfnamefont {R.~S.~K.}\ \bibnamefont
  {Mong}}, \ and\ \bibinfo {author} {\bibfnamefont {F.}~\bibnamefont
  {Pollmann}},\ }\href {\doibase 10.1103/PhysRevLett.110.236801} {\bibfield
  {journal} {\bibinfo  {journal} {Phys. Rev. Lett.}\ }\textbf {\bibinfo
  {volume} {110}},\ \bibinfo {pages} {236801} (\bibinfo {year}
  {2013})}\BibitemShut {NoStop}%
\bibitem [{\citenamefont {Kim}\ \emph {et~al.}(2022{\natexlab{a}})\citenamefont
  {Kim}, \citenamefont {Shi}, \citenamefont {Kato},\ and\ \citenamefont
  {Albert}}]{Kim2022cminus}%
  \BibitemOpen
  \bibfield  {author} {\bibinfo {author} {\bibfnamefont {I.~H.}\ \bibnamefont
  {Kim}}, \bibinfo {author} {\bibfnamefont {B.}~\bibnamefont {Shi}}, \bibinfo
  {author} {\bibfnamefont {K.}~\bibnamefont {Kato}}, \ and\ \bibinfo {author}
  {\bibfnamefont {V.~V.}\ \bibnamefont {Albert}},\ }\href {\doibase
  10.1103/physrevlett.128.176402} {\bibfield  {journal} {\bibinfo  {journal}
  {Physical Review Letters}\ }\textbf {\bibinfo {volume} {128}} (\bibinfo
  {year} {2022}{\natexlab{a}}),\ 10.1103/physrevlett.128.176402}\BibitemShut
  {NoStop}%
\bibitem [{\citenamefont {Kim}\ \emph {et~al.}(2022{\natexlab{b}})\citenamefont
  {Kim}, \citenamefont {Shi}, \citenamefont {Kato},\ and\ \citenamefont
  {Albert}}]{Kim2022modular}%
  \BibitemOpen
  \bibfield  {author} {\bibinfo {author} {\bibfnamefont {I.~H.}\ \bibnamefont
  {Kim}}, \bibinfo {author} {\bibfnamefont {B.}~\bibnamefont {Shi}}, \bibinfo
  {author} {\bibfnamefont {K.}~\bibnamefont {Kato}}, \ and\ \bibinfo {author}
  {\bibfnamefont {V.~V.}\ \bibnamefont {Albert}},\ }\href {\doibase
  10.1103/physrevb.106.075147} {\bibfield  {journal} {\bibinfo  {journal}
  {Physical Review B}\ }\textbf {\bibinfo {volume} {106}} (\bibinfo {year}
  {2022}{\natexlab{b}}),\ 10.1103/physrevb.106.075147}\BibitemShut {NoStop}%
\bibitem [{\citenamefont {Zou}\ \emph {et~al.}(2022)\citenamefont {Zou},
  \citenamefont {Shi}, \citenamefont {Sorce}, \citenamefont {Lim},\ and\
  \citenamefont {Kim}}]{Zou2022modular}%
  \BibitemOpen
  \bibfield  {author} {\bibinfo {author} {\bibfnamefont {Y.}~\bibnamefont
  {Zou}}, \bibinfo {author} {\bibfnamefont {B.}~\bibnamefont {Shi}}, \bibinfo
  {author} {\bibfnamefont {J.}~\bibnamefont {Sorce}}, \bibinfo {author}
  {\bibfnamefont {I.~T.}\ \bibnamefont {Lim}}, \ and\ \bibinfo {author}
  {\bibfnamefont {I.~H.}\ \bibnamefont {Kim}},\ }\href {\doibase
  10.1103/physrevlett.129.260402} {\bibfield  {journal} {\bibinfo  {journal}
  {Physical Review Letters}\ }\textbf {\bibinfo {volume} {129}} (\bibinfo
  {year} {2022}),\ 10.1103/physrevlett.129.260402}\BibitemShut {NoStop}%
\bibitem [{\citenamefont {Fan}(2022)}]{Fan2022cminus}%
  \BibitemOpen
  \bibfield  {author} {\bibinfo {author} {\bibfnamefont {R.}~\bibnamefont
  {Fan}},\ }\href {\doibase 10.1103/physrevlett.129.260403} {\bibfield
  {journal} {\bibinfo  {journal} {Physical Review Letters}\ }\textbf {\bibinfo
  {volume} {129}} (\bibinfo {year} {2022}),\
  10.1103/physrevlett.129.260403}\BibitemShut {NoStop}%
\bibitem [{\citenamefont {Fan}\ \emph {et~al.}(2022)\citenamefont {Fan},
  \citenamefont {Sahay},\ and\ \citenamefont {Vishwanath}}]{Fan2022QHE}%
  \BibitemOpen
  \bibfield  {author} {\bibinfo {author} {\bibfnamefont {R.}~\bibnamefont
  {Fan}}, \bibinfo {author} {\bibfnamefont {R.}~\bibnamefont {Sahay}}, \ and\
  \bibinfo {author} {\bibfnamefont {A.}~\bibnamefont {Vishwanath}},\
  }\href@noop {} {\enquote {\bibinfo {title} {Extracting the quantum hall
  conductance from a single bulk wavefunction},}\ } (\bibinfo {year} {2022}),\
  \Eprint {http://arxiv.org/abs/2208.11710} {arXiv:2208.11710
  [cond-mat.str-el]} \BibitemShut {NoStop}%
\bibitem [{\citenamefont {Shiozaki}\ \emph {et~al.}(2018)\citenamefont
  {Shiozaki}, \citenamefont {Shapourian}, \citenamefont {Gomi},\ and\
  \citenamefont {Ryu}}]{Shiozaki2018antiunitary}%
  \BibitemOpen
  \bibfield  {author} {\bibinfo {author} {\bibfnamefont {K.}~\bibnamefont
  {Shiozaki}}, \bibinfo {author} {\bibfnamefont {H.}~\bibnamefont
  {Shapourian}}, \bibinfo {author} {\bibfnamefont {K.}~\bibnamefont {Gomi}}, \
  and\ \bibinfo {author} {\bibfnamefont {S.}~\bibnamefont {Ryu}},\ }\href
  {\doibase 10.1103/physrevb.98.035151} {\bibfield  {journal} {\bibinfo
  {journal} {Physical Review B}\ }\textbf {\bibinfo {volume} {98}} (\bibinfo
  {year} {2018}),\ 10.1103/physrevb.98.035151}\BibitemShut {NoStop}%
\bibitem [{\citenamefont {Dehghani}\ \emph {et~al.}(2021)\citenamefont
  {Dehghani}, \citenamefont {Cian}, \citenamefont {Hafezi},\ and\ \citenamefont
  {Barkeshli}}]{Dehghani_2021}%
  \BibitemOpen
  \bibfield  {author} {\bibinfo {author} {\bibfnamefont {H.}~\bibnamefont
  {Dehghani}}, \bibinfo {author} {\bibfnamefont {Z.-P.}\ \bibnamefont {Cian}},
  \bibinfo {author} {\bibfnamefont {M.}~\bibnamefont {Hafezi}}, \ and\ \bibinfo
  {author} {\bibfnamefont {M.}~\bibnamefont {Barkeshli}},\ }\href {\doibase
  10.1103/physrevb.103.075102} {\bibfield  {journal} {\bibinfo  {journal}
  {Physical Review B}\ }\textbf {\bibinfo {volume} {103}} (\bibinfo {year}
  {2021}),\ 10.1103/physrevb.103.075102}\BibitemShut {NoStop}%
\bibitem [{\citenamefont {Cian}\ \emph {et~al.}(2021)\citenamefont {Cian},
  \citenamefont {Dehghani}, \citenamefont {Elben}, \citenamefont {Vermersch},
  \citenamefont {Zhu}, \citenamefont {Barkeshli}, \citenamefont {Zoller},\ and\
  \citenamefont {Hafezi}}]{Cian_2021}%
  \BibitemOpen
  \bibfield  {author} {\bibinfo {author} {\bibfnamefont {Z.-P.}\ \bibnamefont
  {Cian}}, \bibinfo {author} {\bibfnamefont {H.}~\bibnamefont {Dehghani}},
  \bibinfo {author} {\bibfnamefont {A.}~\bibnamefont {Elben}}, \bibinfo
  {author} {\bibfnamefont {B.}~\bibnamefont {Vermersch}}, \bibinfo {author}
  {\bibfnamefont {G.}~\bibnamefont {Zhu}}, \bibinfo {author} {\bibfnamefont
  {M.}~\bibnamefont {Barkeshli}}, \bibinfo {author} {\bibfnamefont
  {P.}~\bibnamefont {Zoller}}, \ and\ \bibinfo {author} {\bibfnamefont
  {M.}~\bibnamefont {Hafezi}},\ }\href {\doibase
  10.1103/physrevlett.126.050501} {\bibfield  {journal} {\bibinfo  {journal}
  {Physical Review Letters}\ }\textbf {\bibinfo {volume} {126}} (\bibinfo
  {year} {2021}),\ 10.1103/physrevlett.126.050501}\BibitemShut {NoStop}%
\bibitem [{\citenamefont {Liang}\ \emph
  {et~al.}(2024{\natexlab{a}})\citenamefont {Liang}, \citenamefont {Xu},
  \citenamefont {Iosue},\ and\ \citenamefont {Chen}}]{liang2024extracting}%
  \BibitemOpen
  \bibfield  {author} {\bibinfo {author} {\bibfnamefont {Z.}~\bibnamefont
  {Liang}}, \bibinfo {author} {\bibfnamefont {Y.}~\bibnamefont {Xu}}, \bibinfo
  {author} {\bibfnamefont {J.~T.}\ \bibnamefont {Iosue}}, \ and\ \bibinfo
  {author} {\bibfnamefont {Y.-A.}\ \bibnamefont {Chen}},\ }\href {\doibase
  10.1103/PRXQuantum.5.030328} {\bibfield  {journal} {\bibinfo  {journal} {PRX
  Quantum}\ }\textbf {\bibinfo {volume} {5}},\ \bibinfo {pages} {030328}
  (\bibinfo {year} {2024}{\natexlab{a}})}\BibitemShut {NoStop}%
\bibitem [{\citenamefont {Liang}\ \emph
  {et~al.}(2024{\natexlab{b}})\citenamefont {Liang}, \citenamefont {Yang},
  \citenamefont {Iosue},\ and\ \citenamefont {Chen}}]{liang2024operator}%
  \BibitemOpen
  \bibfield  {author} {\bibinfo {author} {\bibfnamefont {Z.}~\bibnamefont
  {Liang}}, \bibinfo {author} {\bibfnamefont {B.}~\bibnamefont {Yang}},
  \bibinfo {author} {\bibfnamefont {J.~T.}\ \bibnamefont {Iosue}}, \ and\
  \bibinfo {author} {\bibfnamefont {Y.-A.}\ \bibnamefont {Chen}},\ }\href
  {https://arxiv.org/abs/2410.11942} {\bibfield  {journal} {\bibinfo  {journal}
  {arXiv preprint arXiv:2410.11942}\ } (\bibinfo {year}
  {2024}{\natexlab{b}})}\BibitemShut {NoStop}%
\bibitem [{\citenamefont {Kobayashi}\ \emph
  {et~al.}(2024{\natexlab{a}})\citenamefont {Kobayashi}, \citenamefont {Li},
  \citenamefont {Xue}, \citenamefont {Hsin},\ and\ \citenamefont
  {Chen}}]{kobayashi2024universal}%
  \BibitemOpen
  \bibfield  {author} {\bibinfo {author} {\bibfnamefont {R.}~\bibnamefont
  {Kobayashi}}, \bibinfo {author} {\bibfnamefont {Y.}~\bibnamefont {Li}},
  \bibinfo {author} {\bibfnamefont {H.}~\bibnamefont {Xue}}, \bibinfo {author}
  {\bibfnamefont {P.-S.}\ \bibnamefont {Hsin}}, \ and\ \bibinfo {author}
  {\bibfnamefont {Y.-A.}\ \bibnamefont {Chen}},\ }\href
  {https://arxiv.org/abs/2412.01886} {\bibfield  {journal} {\bibinfo  {journal}
  {arXiv preprint arXiv:2412.01886}\ } (\bibinfo {year}
  {2024}{\natexlab{a}})}\BibitemShut {NoStop}%
\bibitem [{\citenamefont {Kapustin}\ and\ \citenamefont
  {Saulina}(2011)}]{Kapustin:2010hk}%
  \BibitemOpen
  \bibfield  {author} {\bibinfo {author} {\bibfnamefont {A.}~\bibnamefont
  {Kapustin}}\ and\ \bibinfo {author} {\bibfnamefont {N.}~\bibnamefont
  {Saulina}},\ }\href {\doibase 10.1016/j.nuclphysb.2010.12.017} {\bibfield
  {journal} {\bibinfo  {journal} {Nucl. Phys.}\ }\textbf {\bibinfo {volume}
  {B845}},\ \bibinfo {pages} {393} (\bibinfo {year} {2011})},\ \Eprint
  {http://arxiv.org/abs/1008.0654} {arXiv:1008.0654 [hep-th]} \BibitemShut
  {NoStop}%
\bibitem [{\citenamefont {Levin}(2013)}]{Levin2013edge}%
  \BibitemOpen
  \bibfield  {author} {\bibinfo {author} {\bibfnamefont {M.}~\bibnamefont
  {Levin}},\ }\href {\doibase 10.1103/physrevx.3.021009} {\bibfield  {journal}
  {\bibinfo  {journal} {Physical Review X}\ }\textbf {\bibinfo {volume} {3}}
  (\bibinfo {year} {2013}),\ 10.1103/physrevx.3.021009},\ \Eprint
  {http://arxiv.org/abs/1301.7355} {arXiv:1301.7355 [cond-mat.str-el]}
  \BibitemShut {NoStop}%
\bibitem [{\citenamefont {Ng}\ \emph {et~al.}(2019)\citenamefont {Ng},
  \citenamefont {Schopieray},\ and\ \citenamefont {Wang}}]{Ng2018higher}%
  \BibitemOpen
  \bibfield  {author} {\bibinfo {author} {\bibfnamefont {S.-H.}\ \bibnamefont
  {Ng}}, \bibinfo {author} {\bibfnamefont {A.}~\bibnamefont {Schopieray}}, \
  and\ \bibinfo {author} {\bibfnamefont {Y.}~\bibnamefont {Wang}},\ }\href
  {\doibase 10.1007/s00029-019-0499-2} {\bibfield  {journal} {\bibinfo
  {journal} {Selecta Mathematica}\ }\textbf {\bibinfo {volume} {25}} (\bibinfo
  {year} {2019}),\ 10.1007/s00029-019-0499-2},\ \Eprint
  {http://arxiv.org/abs/1812.11234} {arXiv:1812.11234 [math.QA]} \BibitemShut
  {NoStop}%
\bibitem [{\citenamefont {Ng}\ \emph {et~al.}(2020)\citenamefont {Ng},
  \citenamefont {Rowell}, \citenamefont {Wang},\ and\ \citenamefont
  {Zhang}}]{Ng2020higher}%
  \BibitemOpen
  \bibfield  {author} {\bibinfo {author} {\bibfnamefont {S.-H.}\ \bibnamefont
  {Ng}}, \bibinfo {author} {\bibfnamefont {E.~C.}\ \bibnamefont {Rowell}},
  \bibinfo {author} {\bibfnamefont {Y.}~\bibnamefont {Wang}}, \ and\ \bibinfo
  {author} {\bibfnamefont {Q.}~\bibnamefont {Zhang}},\ }\href
  {https://arxiv.org/abs/2002.03570} {\enquote {\bibinfo {title} {Higher
  central charges and witt groups},}\ } (\bibinfo {year} {2020}),\ \Eprint
  {http://arxiv.org/abs/2002.03570} {arXiv:2002.03570 [math.QA]} \BibitemShut
  {NoStop}%
\bibitem [{\citenamefont {Kaidi}\ \emph {et~al.}(2021)\citenamefont {Kaidi},
  \citenamefont {Komargodski}, \citenamefont {Ohmori}, \citenamefont
  {Seifnashri},\ and\ \citenamefont {Shao}}]{kaidi2021higher}%
  \BibitemOpen
  \bibfield  {author} {\bibinfo {author} {\bibfnamefont {J.}~\bibnamefont
  {Kaidi}}, \bibinfo {author} {\bibfnamefont {Z.}~\bibnamefont {Komargodski}},
  \bibinfo {author} {\bibfnamefont {K.}~\bibnamefont {Ohmori}}, \bibinfo
  {author} {\bibfnamefont {S.}~\bibnamefont {Seifnashri}}, \ and\ \bibinfo
  {author} {\bibfnamefont {S.-H.}\ \bibnamefont {Shao}},\ }\href
  {https://arxiv.org/abs/2107.13091} {\enquote {\bibinfo {title} {Higher
  central charges and topological boundaries in 2+1-dimensional tqfts},}\ }
  (\bibinfo {year} {2021}),\ \Eprint {http://arxiv.org/abs/2107.13091}
  {arXiv:2107.13091 [hep-th]} \BibitemShut {NoStop}%
\bibitem [{\citenamefont {Kobayashi}(2022)}]{Kobayashi2022FQH}%
  \BibitemOpen
  \bibfield  {author} {\bibinfo {author} {\bibfnamefont {R.}~\bibnamefont
  {Kobayashi}},\ }\href {\doibase 10.1103/physrevresearch.4.033137} {\bibfield
  {journal} {\bibinfo  {journal} {Physical Review Research}\ }\textbf {\bibinfo
  {volume} {4}} (\bibinfo {year} {2022}),\ 10.1103/physrevresearch.4.033137},\
  \Eprint {http://arxiv.org/abs/arXiv:2203.08156} {arXiv:2203.08156}
  \BibitemShut {NoStop}%
\bibitem [{\citenamefont {Kobayashi}\ \emph
  {et~al.}(2024{\natexlab{b}})\citenamefont {Kobayashi}, \citenamefont {Wang},
  \citenamefont {Soejima}, \citenamefont {Mong},\ and\ \citenamefont
  {Ryu}}]{Kobayashi2024hcc}%
  \BibitemOpen
  \bibfield  {author} {\bibinfo {author} {\bibfnamefont {R.}~\bibnamefont
  {Kobayashi}}, \bibinfo {author} {\bibfnamefont {T.}~\bibnamefont {Wang}},
  \bibinfo {author} {\bibfnamefont {T.}~\bibnamefont {Soejima}}, \bibinfo
  {author} {\bibfnamefont {R.~S.~K.}\ \bibnamefont {Mong}}, \ and\ \bibinfo
  {author} {\bibfnamefont {S.}~\bibnamefont {Ryu}},\ }\href {\doibase
  10.1103/physrevlett.132.016602} {\bibfield  {journal} {\bibinfo  {journal}
  {Physical Review Letters}\ }\textbf {\bibinfo {volume} {132}} (\bibinfo
  {year} {2024}{\natexlab{b}}),\ 10.1103/physrevlett.132.016602}\BibitemShut
  {NoStop}%
\bibitem [{\citenamefont {Cano}\ \emph {et~al.}(2015)\citenamefont {Cano},
  \citenamefont {Hughes},\ and\ \citenamefont
  {Mulligan}}]{Cano2015Interaction}%
  \BibitemOpen
  \bibfield  {author} {\bibinfo {author} {\bibfnamefont {J.}~\bibnamefont
  {Cano}}, \bibinfo {author} {\bibfnamefont {T.~L.}\ \bibnamefont {Hughes}}, \
  and\ \bibinfo {author} {\bibfnamefont {M.}~\bibnamefont {Mulligan}},\ }\href
  {\doibase 10.1103/physrevb.92.075104} {\bibfield  {journal} {\bibinfo
  {journal} {Physical Review B}\ }\textbf {\bibinfo {volume} {92}} (\bibinfo
  {year} {2015}),\ 10.1103/physrevb.92.075104}\BibitemShut {NoStop}%
\bibitem [{\citenamefont {Zou}\ and\ \citenamefont
  {Haah}(2016)}]{Zou2016spurious}%
  \BibitemOpen
  \bibfield  {author} {\bibinfo {author} {\bibfnamefont {L.}~\bibnamefont
  {Zou}}\ and\ \bibinfo {author} {\bibfnamefont {J.}~\bibnamefont {Haah}},\
  }\href {\doibase 10.1103/physrevb.94.075151} {\bibfield  {journal} {\bibinfo
  {journal} {Physical Review B}\ }\textbf {\bibinfo {volume} {94}} (\bibinfo
  {year} {2016}),\ 10.1103/physrevb.94.075151}\BibitemShut {NoStop}%
\bibitem [{\citenamefont {Williamson}\ \emph {et~al.}(2019)\citenamefont
  {Williamson}, \citenamefont {Dua},\ and\ \citenamefont
  {Cheng}}]{Williamson2019spurious}%
  \BibitemOpen
  \bibfield  {author} {\bibinfo {author} {\bibfnamefont {D.~J.}\ \bibnamefont
  {Williamson}}, \bibinfo {author} {\bibfnamefont {A.}~\bibnamefont {Dua}}, \
  and\ \bibinfo {author} {\bibfnamefont {M.}~\bibnamefont {Cheng}},\ }\href
  {\doibase 10.1103/physrevlett.122.140506} {\bibfield  {journal} {\bibinfo
  {journal} {Physical Review Letters}\ }\textbf {\bibinfo {volume} {122}}
  (\bibinfo {year} {2019}),\ 10.1103/physrevlett.122.140506}\BibitemShut
  {NoStop}%
\bibitem [{\citenamefont {Kim}\ \emph {et~al.}(2023)\citenamefont {Kim},
  \citenamefont {Levin}, \citenamefont {Lin}, \citenamefont {Ranard},\ and\
  \citenamefont {Shi}}]{Kim2023bound}%
  \BibitemOpen
  \bibfield  {author} {\bibinfo {author} {\bibfnamefont {I.~H.}\ \bibnamefont
  {Kim}}, \bibinfo {author} {\bibfnamefont {M.}~\bibnamefont {Levin}}, \bibinfo
  {author} {\bibfnamefont {T.-C.}\ \bibnamefont {Lin}}, \bibinfo {author}
  {\bibfnamefont {D.}~\bibnamefont {Ranard}}, \ and\ \bibinfo {author}
  {\bibfnamefont {B.}~\bibnamefont {Shi}},\ }\href {\doibase
  10.1103/physrevlett.131.166601} {\bibfield  {journal} {\bibinfo  {journal}
  {Physical Review Letters}\ }\textbf {\bibinfo {volume} {131}} (\bibinfo
  {year} {2023}),\ 10.1103/physrevlett.131.166601}\BibitemShut {NoStop}%
\bibitem [{\citenamefont {Levin}(2024)}]{Levin2024inequality}%
  \BibitemOpen
  \bibfield  {author} {\bibinfo {author} {\bibfnamefont {M.}~\bibnamefont
  {Levin}},\ }\href {\doibase 10.1103/PhysRevB.110.165154} {\bibfield
  {journal} {\bibinfo  {journal} {Phys. Rev. B}\ }\textbf {\bibinfo {volume}
  {110}},\ \bibinfo {pages} {165154} (\bibinfo {year} {2024})}\BibitemShut
  {NoStop}%
\bibitem [{\citenamefont {Lieb}\ \emph {et~al.}(1961)\citenamefont {Lieb},
  \citenamefont {Schultz},\ and\ \citenamefont {Mattis}}]{Lieb1961}%
  \BibitemOpen
  \bibfield  {author} {\bibinfo {author} {\bibfnamefont {E.}~\bibnamefont
  {Lieb}}, \bibinfo {author} {\bibfnamefont {T.}~\bibnamefont {Schultz}}, \
  and\ \bibinfo {author} {\bibfnamefont {D.}~\bibnamefont {Mattis}},\ }\href
  {\doibase 10.1016/0003-4916(61)90115-4} {\bibfield  {journal} {\bibinfo
  {journal} {Annals of Physics}\ }\textbf {\bibinfo {volume} {16}},\ \bibinfo
  {pages} {407} (\bibinfo {year} {1961})}\BibitemShut {NoStop}%
\bibitem [{\citenamefont {Oshikawa}(2000)}]{Oshikawa}%
  \BibitemOpen
  \bibfield  {author} {\bibinfo {author} {\bibfnamefont {M.}~\bibnamefont
  {Oshikawa}},\ }\href {\doibase 10.1103/PhysRevLett.84.1535} {\bibfield
  {journal} {\bibinfo  {journal} {Physical Review Letters}\ }\textbf {\bibinfo
  {volume} {84}},\ \bibinfo {pages} {1535} (\bibinfo {year} {2000})},\ \Eprint
  {http://arxiv.org/abs/cond-mat/9911137} {arXiv:cond-mat/9911137} \BibitemShut
  {NoStop}%
\bibitem [{\citenamefont {Hastings}(2004)}]{Hastings}%
  \BibitemOpen
  \bibfield  {author} {\bibinfo {author} {\bibfnamefont {M.}~\bibnamefont
  {Hastings}},\ }\href {\doibase 10.1103/PhysRevB.69.104431} {\bibfield
  {journal} {\bibinfo  {journal} {Physical Review B}\ }\textbf {\bibinfo
  {volume} {69}},\ \bibinfo {pages} {104431} (\bibinfo {year} {2004})},\
  \Eprint {http://arxiv.org/abs/cond-mat/0305505} {arXiv:cond-mat/0305505}
  \BibitemShut {NoStop}%
\bibitem [{\citenamefont {Kobayashi}\ and\ \citenamefont
  {Barkeshli}(2023)}]{kobayashi202331d}%
  \BibitemOpen
  \bibfield  {author} {\bibinfo {author} {\bibfnamefont {R.}~\bibnamefont
  {Kobayashi}}\ and\ \bibinfo {author} {\bibfnamefont {M.}~\bibnamefont
  {Barkeshli}},\ }\href@noop {} {\enquote {\bibinfo {title} {(3+1)d path
  integral state sums on curved u(1) bundles and u(1) anomalies of (2+1)d
  topological phases},}\ } (\bibinfo {year} {2023}),\ \Eprint
  {http://arxiv.org/abs/2111.14827} {arXiv:2111.14827 [cond-mat.str-el]}
  \BibitemShut {NoStop}%
\bibitem [{\citenamefont {Lapa}\ and\ \citenamefont {Levin}(2019)}]{Lapa2019}%
  \BibitemOpen
  \bibfield  {author} {\bibinfo {author} {\bibfnamefont {M.~F.}\ \bibnamefont
  {Lapa}}\ and\ \bibinfo {author} {\bibfnamefont {M.}~\bibnamefont {Levin}},\
  }\href {\doibase 10.1103/physrevb.100.165129} {\bibfield  {journal} {\bibinfo
   {journal} {Physical Review B}\ }\textbf {\bibinfo {volume} {100}} (\bibinfo
  {year} {2019}),\ 10.1103/physrevb.100.165129}\BibitemShut {NoStop}%
\bibitem [{\citenamefont {Qi}\ \emph {et~al.}(2012)\citenamefont {Qi},
  \citenamefont {Katsura},\ and\ \citenamefont {Ludwig}}]{Qi2012entanglement}%
  \BibitemOpen
  \bibfield  {author} {\bibinfo {author} {\bibfnamefont {X.-L.}\ \bibnamefont
  {Qi}}, \bibinfo {author} {\bibfnamefont {H.}~\bibnamefont {Katsura}}, \ and\
  \bibinfo {author} {\bibfnamefont {A.~W.~W.}\ \bibnamefont {Ludwig}},\ }\href
  {\doibase 10.1103/physrevlett.108.196402} {\bibfield  {journal} {\bibinfo
  {journal} {Physical Review Letters}\ }\textbf {\bibinfo {volume} {108}}
  (\bibinfo {year} {2012}),\ 10.1103/physrevlett.108.196402},\ \Eprint
  {http://arxiv.org/abs/1103.5437} {arXiv:1103.5437 [cond-mat.mes-hall]}
  \BibitemShut {NoStop}%
\bibitem [{\citenamefont {Kitaev}\ and\ \citenamefont
  {Preskill}(2006)}]{KitaevTEE}%
  \BibitemOpen
  \bibfield  {author} {\bibinfo {author} {\bibfnamefont {A.}~\bibnamefont
  {Kitaev}}\ and\ \bibinfo {author} {\bibfnamefont {J.}~\bibnamefont
  {Preskill}},\ }\href {\doibase 10.1103/physrevlett.96.110404} {\bibfield
  {journal} {\bibinfo  {journal} {Physical Review Letters}\ }\textbf {\bibinfo
  {volume} {96}} (\bibinfo {year} {2006}),\ 10.1103/physrevlett.96.110404},\
  \Eprint {http://arxiv.org/abs/hep-th/0510092} {hep-th/0510092} \BibitemShut
  {NoStop}%
\bibitem [{\citenamefont {Zaletel}\ and\ \citenamefont
  {Mong}(2012)}]{exactmps}%
  \BibitemOpen
  \bibfield  {author} {\bibinfo {author} {\bibfnamefont {M.~P.}\ \bibnamefont
  {Zaletel}}\ and\ \bibinfo {author} {\bibfnamefont {R.~S.~K.}\ \bibnamefont
  {Mong}},\ }\href {\doibase 10.1103/PhysRevB.86.245305} {\bibfield  {journal}
  {\bibinfo  {journal} {Phys. Rev. B}\ }\textbf {\bibinfo {volume} {86}},\
  \bibinfo {pages} {245305} (\bibinfo {year} {2012})}\BibitemShut {NoStop}%
\bibitem [{\citenamefont {Rezayi}\ and\ \citenamefont
  {Haldane}(1994)}]{Haldane_1994_pseudopotential}%
  \BibitemOpen
  \bibfield  {author} {\bibinfo {author} {\bibfnamefont {E.~H.}\ \bibnamefont
  {Rezayi}}\ and\ \bibinfo {author} {\bibfnamefont {F.~D.~M.}\ \bibnamefont
  {Haldane}},\ }\href {\doibase 10.1103/PhysRevB.50.17199} {\bibfield
  {journal} {\bibinfo  {journal} {Phys. Rev. B}\ }\textbf {\bibinfo {volume}
  {50}},\ \bibinfo {pages} {17199} (\bibinfo {year} {1994})}\BibitemShut
  {NoStop}%
\bibitem [{\citenamefont {Halperin}\ \emph {et~al.}(2020)\citenamefont
  {Halperin}, \citenamefont {Jain},\ and\ \citenamefont
  {Cooper}}]{halperin_jain_cooper_2020}%
  \BibitemOpen
  \bibfield  {author} {\bibinfo {author} {\bibfnamefont {B.~I.}\ \bibnamefont
  {Halperin}}, \bibinfo {author} {\bibfnamefont {J.~K.}\ \bibnamefont {Jain}},
  \ and\ \bibinfo {author} {\bibfnamefont {N.~R.}\ \bibnamefont {Cooper}},\
  }\enquote {\bibinfo {title} {Fractional quantum hall states of bosons:
  Properties and prospects for experimental realization},}\ in\ \href@noop {}
  {\emph {\bibinfo {booktitle} {Fractional quantum hall effects: New
  developments}}}\ (\bibinfo  {publisher} {World Scientific},\ \bibinfo {year}
  {2020})\ p.\ \bibinfo {pages} {487–521}\BibitemShut {NoStop}%
\bibitem [{\citenamefont {Zaletel}\ \emph {et~al.}(2015)\citenamefont
  {Zaletel}, \citenamefont {Mong}, \citenamefont {Pollmann},\ and\
  \citenamefont {Rezayi}}]{PhysRevB.91.045115}%
  \BibitemOpen
  \bibfield  {author} {\bibinfo {author} {\bibfnamefont {M.~P.}\ \bibnamefont
  {Zaletel}}, \bibinfo {author} {\bibfnamefont {R.~S.~K.}\ \bibnamefont
  {Mong}}, \bibinfo {author} {\bibfnamefont {F.}~\bibnamefont {Pollmann}}, \
  and\ \bibinfo {author} {\bibfnamefont {E.~H.}\ \bibnamefont {Rezayi}},\
  }\href {\doibase 10.1103/PhysRevB.91.045115} {\bibfield  {journal} {\bibinfo
  {journal} {Phys. Rev. B}\ }\textbf {\bibinfo {volume} {91}},\ \bibinfo
  {pages} {045115} (\bibinfo {year} {2015})}\BibitemShut {NoStop}%
\bibitem [{\citenamefont {Bruillard}\ \emph {et~al.}(2015)\citenamefont
  {Bruillard}, \citenamefont {Ng}, \citenamefont {Rowell},\ and\ \citenamefont
  {Wang}}]{Bruillard2015rankfiniteness}%
  \BibitemOpen
  \bibfield  {author} {\bibinfo {author} {\bibfnamefont {P.}~\bibnamefont
  {Bruillard}}, \bibinfo {author} {\bibfnamefont {S.-H.}\ \bibnamefont {Ng}},
  \bibinfo {author} {\bibfnamefont {E.}~\bibnamefont {Rowell}}, \ and\ \bibinfo
  {author} {\bibfnamefont {Z.}~\bibnamefont {Wang}},\ }\href {\doibase
  10.1090/jams/842} {\bibfield  {journal} {\bibinfo  {journal} {Journal of the
  American Mathematical Society}\ }\textbf {\bibinfo {volume} {29}},\ \bibinfo
  {pages} {857} (\bibinfo {year} {2015})},\ \Eprint
  {http://arxiv.org/abs/arXiv:1310.7050} {arXiv:1310.7050} \BibitemShut
  {NoStop}%
\bibitem [{\citenamefont {Barkeshli}\ \emph {et~al.}(2019)\citenamefont
  {Barkeshli}, \citenamefont {Bonderson}, \citenamefont {Cheng},\ and\
  \citenamefont {Wang}}]{barkeshli2019}%
  \BibitemOpen
  \bibfield  {author} {\bibinfo {author} {\bibfnamefont {M.}~\bibnamefont
  {Barkeshli}}, \bibinfo {author} {\bibfnamefont {P.}~\bibnamefont
  {Bonderson}}, \bibinfo {author} {\bibfnamefont {M.}~\bibnamefont {Cheng}}, \
  and\ \bibinfo {author} {\bibfnamefont {Z.}~\bibnamefont {Wang}},\ }\href
  {\doibase 10.1103/PhysRevB.100.115147} {\bibfield  {journal} {\bibinfo
  {journal} {Phys. Rev. B}\ }\textbf {\bibinfo {volume} {100}},\ \bibinfo
  {pages} {115147} (\bibinfo {year} {2019})},\ \Eprint
  {http://arxiv.org/abs/arXiv:1410.4540} {arXiv:1410.4540} \BibitemShut
  {NoStop}%
\bibitem [{\citenamefont {Bulmash}\ and\ \citenamefont
  {Barkeshli}(2022{\natexlab{a}})}]{Bulmash2022fermfrac}%
  \BibitemOpen
  \bibfield  {author} {\bibinfo {author} {\bibfnamefont {D.}~\bibnamefont
  {Bulmash}}\ and\ \bibinfo {author} {\bibfnamefont {M.}~\bibnamefont
  {Barkeshli}},\ }\href {\doibase 10.1103/physrevb.105.125114} {\bibfield
  {journal} {\bibinfo  {journal} {Physical Review B}\ }\textbf {\bibinfo
  {volume} {105}} (\bibinfo {year} {2022}{\natexlab{a}}),\
  10.1103/physrevb.105.125114}\BibitemShut {NoStop}%
\bibitem [{\citenamefont {Benini}\ \emph {et~al.}(2019)\citenamefont {Benini},
  \citenamefont {Córdova},\ and\ \citenamefont {Hsin}}]{Benini2019_2group}%
  \BibitemOpen
  \bibfield  {author} {\bibinfo {author} {\bibfnamefont {F.}~\bibnamefont
  {Benini}}, \bibinfo {author} {\bibfnamefont {C.}~\bibnamefont {Córdova}}, \
  and\ \bibinfo {author} {\bibfnamefont {P.-S.}\ \bibnamefont {Hsin}},\ }\href
  {\doibase 10.1007/jhep03(2019)118} {\bibfield  {journal} {\bibinfo  {journal}
  {Journal of High Energy Physics}\ }\textbf {\bibinfo {volume} {2019}}
  (\bibinfo {year} {2019}),\ 10.1007/jhep03(2019)118}\BibitemShut {NoStop}%
\bibitem [{\citenamefont {Cheng}\ \emph {et~al.}(2023)\citenamefont {Cheng},
  \citenamefont {Hsin},\ and\ \citenamefont {Jian}}]{Cheng2023gauging}%
  \BibitemOpen
  \bibfield  {author} {\bibinfo {author} {\bibfnamefont {M.}~\bibnamefont
  {Cheng}}, \bibinfo {author} {\bibfnamefont {P.-S.}\ \bibnamefont {Hsin}}, \
  and\ \bibinfo {author} {\bibfnamefont {C.-M.}\ \bibnamefont {Jian}},\ }\href
  {\doibase 10.21468/scipostphys.14.5.100} {\bibfield  {journal} {\bibinfo
  {journal} {SciPost Physics}\ }\textbf {\bibinfo {volume} {14}} (\bibinfo
  {year} {2023}),\ 10.21468/scipostphys.14.5.100}\BibitemShut {NoStop}%
\bibitem [{\citenamefont {Thorngren}(2019)}]{thorngren2019anomalies}%
  \BibitemOpen
  \bibfield  {author} {\bibinfo {author} {\bibfnamefont {R.}~\bibnamefont
  {Thorngren}},\ }\href@noop {} {\enquote {\bibinfo {title} {Anomalies and
  bosonization},}\ } (\bibinfo {year} {2019}),\ \Eprint
  {http://arxiv.org/abs/1810.04414} {arXiv:1810.04414 [cond-mat.str-el]}
  \BibitemShut {NoStop}%
\bibitem [{\citenamefont {Thorngren}\ and\ \citenamefont
  {Wang}(2021)}]{Thorngren2021end}%
  \BibitemOpen
  \bibfield  {author} {\bibinfo {author} {\bibfnamefont {R.}~\bibnamefont
  {Thorngren}}\ and\ \bibinfo {author} {\bibfnamefont {Y.}~\bibnamefont
  {Wang}},\ }\href {\doibase 10.1007/jhep09(2021)017} {\bibfield  {journal}
  {\bibinfo  {journal} {Journal of High Energy Physics}\ }\textbf {\bibinfo
  {volume} {2021}} (\bibinfo {year} {2021}),\
  10.1007/jhep09(2021)017}\BibitemShut {NoStop}%
\bibitem [{\citenamefont {Manjunath}\ and\ \citenamefont
  {Barkeshli}(2020)}]{manjunath2020FQH}%
  \BibitemOpen
  \bibfield  {author} {\bibinfo {author} {\bibfnamefont {N.}~\bibnamefont
  {Manjunath}}\ and\ \bibinfo {author} {\bibfnamefont {M.}~\bibnamefont
  {Barkeshli}},\ }\href {\doibase 10.48550/ARXIV.2012.11603} {\enquote
  {\bibinfo {title} {Classification of fractional quantum hall states with
  spatial symmetries},}\ } (\bibinfo {year} {2020}),\ \Eprint
  {http://arxiv.org/abs/2012.11603} {arXiv:2012.11603 [cond-mat.str-el]}
  \BibitemShut {NoStop}%
\bibitem [{\citenamefont {Neupert}\ \emph {et~al.}(2016)\citenamefont
  {Neupert}, \citenamefont {He}, \citenamefont {Keyserlingk}, \citenamefont
  {Sierra},\ and\ \citenamefont {Bernevig}}]{Neupert2016nogo}%
  \BibitemOpen
  \bibfield  {author} {\bibinfo {author} {\bibfnamefont {T.}~\bibnamefont
  {Neupert}}, \bibinfo {author} {\bibfnamefont {H.}~\bibnamefont {He}},
  \bibinfo {author} {\bibfnamefont {C.~v.}\ \bibnamefont {Keyserlingk}},
  \bibinfo {author} {\bibfnamefont {G.}~\bibnamefont {Sierra}}, \ and\ \bibinfo
  {author} {\bibfnamefont {B.~A.}\ \bibnamefont {Bernevig}},\ }\href {\doibase
  10.1088/1367-2630/18/12/123009} {\bibfield  {journal} {\bibinfo  {journal}
  {New Journal of Physics}\ }\textbf {\bibinfo {volume} {18}},\ \bibinfo
  {pages} {123009} (\bibinfo {year} {2016})}\BibitemShut {NoStop}%
\bibitem [{\citenamefont {You}(2023)}]{you2023gapped}%
  \BibitemOpen
  \bibfield  {author} {\bibinfo {author} {\bibfnamefont {M.}~\bibnamefont
  {You}},\ }\href@noop {} {\enquote {\bibinfo {title} {Gapped boundaries of
  fermionic topological orders and higher central charges},}\ } (\bibinfo
  {year} {2023}),\ \Eprint {http://arxiv.org/abs/2311.01096} {arXiv:2311.01096
  [cond-mat.str-el]} \BibitemShut {NoStop}%
\bibitem [{\citenamefont {Thorngren}\ and\ \citenamefont
  {Wang}(2019)}]{thorngren2019fusioncategorysymmetryi}%
  \BibitemOpen
  \bibfield  {author} {\bibinfo {author} {\bibfnamefont {R.}~\bibnamefont
  {Thorngren}}\ and\ \bibinfo {author} {\bibfnamefont {Y.}~\bibnamefont
  {Wang}},\ }\href {https://arxiv.org/abs/1912.02817} {\enquote {\bibinfo
  {title} {Fusion category symmetry i: Anomaly in-flow and gapped phases},}\ }
  (\bibinfo {year} {2019}),\ \Eprint {http://arxiv.org/abs/1912.02817}
  {arXiv:1912.02817 [hep-th]} \BibitemShut {NoStop}%
\bibitem [{\citenamefont {Zhang}\ and\ \citenamefont
  {C\'ordova}(2023)}]{Zhang:2023wlu}%
  \BibitemOpen
  \bibfield  {author} {\bibinfo {author} {\bibfnamefont {C.}~\bibnamefont
  {Zhang}}\ and\ \bibinfo {author} {\bibfnamefont {C.}~\bibnamefont
  {C\'ordova}},\ }\href@noop {} {\  (\bibinfo {year} {2023})},\ \Eprint
  {http://arxiv.org/abs/2304.01262} {arXiv:2304.01262 [cond-mat.str-el]}
  \BibitemShut {NoStop}%
\bibitem [{\citenamefont {Cordova}\ \emph {et~al.}(2023)\citenamefont
  {Cordova}, \citenamefont {Hsin},\ and\ \citenamefont
  {Zhang}}]{Cordova:2023bja}%
  \BibitemOpen
  \bibfield  {author} {\bibinfo {author} {\bibfnamefont {C.}~\bibnamefont
  {Cordova}}, \bibinfo {author} {\bibfnamefont {P.-S.}\ \bibnamefont {Hsin}}, \
  and\ \bibinfo {author} {\bibfnamefont {C.}~\bibnamefont {Zhang}},\
  }\href@noop {} {\  (\bibinfo {year} {2023})},\ \Eprint
  {http://arxiv.org/abs/2308.11706} {arXiv:2308.11706 [hep-th]} \BibitemShut
  {NoStop}%
\bibitem [{\citenamefont {Hsin}\ \emph {et~al.}(2024)\citenamefont {Hsin},
  \citenamefont {Kobayashi},\ and\ \citenamefont {Zhang}}]{Hsin2024coset}%
  \BibitemOpen
  \bibfield  {author} {\bibinfo {author} {\bibfnamefont {P.-S.}\ \bibnamefont
  {Hsin}}, \bibinfo {author} {\bibfnamefont {R.}~\bibnamefont {Kobayashi}}, \
  and\ \bibinfo {author} {\bibfnamefont {C.}~\bibnamefont {Zhang}},\ }\href
  {\doibase 10.21468/scipostphys.17.3.095} {\bibfield  {journal} {\bibinfo
  {journal} {SciPost Physics}\ }\textbf {\bibinfo {volume} {17}} (\bibinfo
  {year} {2024}),\ 10.21468/scipostphys.17.3.095}\BibitemShut {NoStop}%
\bibitem [{\citenamefont {Shiozaki}\ \emph {et~al.}(2017)\citenamefont
  {Shiozaki}, \citenamefont {Shapourian},\ and\ \citenamefont
  {Ryu}}]{Shiozaki2017point}%
  \BibitemOpen
  \bibfield  {author} {\bibinfo {author} {\bibfnamefont {K.}~\bibnamefont
  {Shiozaki}}, \bibinfo {author} {\bibfnamefont {H.}~\bibnamefont
  {Shapourian}}, \ and\ \bibinfo {author} {\bibfnamefont {S.}~\bibnamefont
  {Ryu}},\ }\href {\doibase 10.1103/physrevb.95.205139} {\bibfield  {journal}
  {\bibinfo  {journal} {Physical Review B}\ }\textbf {\bibinfo {volume} {95}}
  (\bibinfo {year} {2017}),\ 10.1103/physrevb.95.205139},\ \Eprint
  {http://arxiv.org/abs/1609.05970} {arXiv:1609.05970 [cond-mat.str-el]}
  \BibitemShut {NoStop}%
\bibitem [{\citenamefont {Zhang}\ \emph {et~al.}(2023)\citenamefont {Zhang},
  \citenamefont {Manjunath}, \citenamefont {Kobayashi},\ and\ \citenamefont
  {Barkeshli}}]{Zhang2023partial}%
  \BibitemOpen
  \bibfield  {author} {\bibinfo {author} {\bibfnamefont {Y.}~\bibnamefont
  {Zhang}}, \bibinfo {author} {\bibfnamefont {N.}~\bibnamefont {Manjunath}},
  \bibinfo {author} {\bibfnamefont {R.}~\bibnamefont {Kobayashi}}, \ and\
  \bibinfo {author} {\bibfnamefont {M.}~\bibnamefont {Barkeshli}},\ }\href
  {\doibase 10.1103/physrevlett.131.176501} {\bibfield  {journal} {\bibinfo
  {journal} {Physical Review Letters}\ }\textbf {\bibinfo {volume} {131}}
  (\bibinfo {year} {2023}),\ 10.1103/physrevlett.131.176501}\BibitemShut
  {NoStop}%
\bibitem [{\citenamefont {Kobayashi}\ \emph
  {et~al.}(2024{\natexlab{c}})\citenamefont {Kobayashi}, \citenamefont {Zhang},
  \citenamefont {Wang},\ and\ \citenamefont {Barkeshli}}]{kobayashi2024RT}%
  \BibitemOpen
  \bibfield  {author} {\bibinfo {author} {\bibfnamefont {R.}~\bibnamefont
  {Kobayashi}}, \bibinfo {author} {\bibfnamefont {Y.}~\bibnamefont {Zhang}},
  \bibinfo {author} {\bibfnamefont {Y.-Q.}\ \bibnamefont {Wang}}, \ and\
  \bibinfo {author} {\bibfnamefont {M.}~\bibnamefont {Barkeshli}},\ }\href@noop
  {} {\enquote {\bibinfo {title} {(2+1)d topological phases with rt symmetry:
  many-body invariant, classification, and higher order edge modes},}\ }
  (\bibinfo {year} {2024}{\natexlab{c}}),\ \Eprint
  {http://arxiv.org/abs/2403.18887} {arXiv:2403.18887 [cond-mat.str-el]}
  \BibitemShut {NoStop}%
\bibitem [{\citenamefont {Bulmash}\ and\ \citenamefont
  {Barkeshli}(2022{\natexlab{b}})}]{Bulmash2022fermionic}%
  \BibitemOpen
  \bibfield  {author} {\bibinfo {author} {\bibfnamefont {D.}~\bibnamefont
  {Bulmash}}\ and\ \bibinfo {author} {\bibfnamefont {M.}~\bibnamefont
  {Barkeshli}},\ }\href {\doibase 10.1103/physrevb.105.125114} {\bibfield
  {journal} {\bibinfo  {journal} {Physical Review B}\ }\textbf {\bibinfo
  {volume} {105}} (\bibinfo {year} {2022}{\natexlab{b}}),\
  10.1103/physrevb.105.125114},\ \Eprint {http://arxiv.org/abs/2109.10913}
  {arXiv:2109.10913 [cond-mat.str-el]} \BibitemShut {NoStop}%
\end{thebibliography}%
\clearpage
\appendix

\begin{widetext}

\section{Detailed computations of the partial rotation}
\label{app:detailedCFT}
\subsection{Fermionic case}
Here we compute the phase of the quantity given by
\begin{align}
     \mathcal{T}^f_a\left(\frac{2\pi}{n},\frac{\pi}{n}\right):=\bra{\Psi_a}U_{\frac{\pi}{n},\mathrm{A}}T_{\frac{2\pi}{n},\mathrm{A}} \ket{\Psi_a}
\end{align}
in terms of edge CFT with $\U^f$ symmetry by utilizing the cut-and-glue approach~\cite{Qi2012entanglement}. For simplicity, let us consider the case where the edge CFT is described in terms of chiral rational CFT; the generalization to non-chiral case is rather straightforward, see Ref.~\onlinecite{Kobayashi2024hcc} for the discussions in bosonic non-chiral phases.

This approach describes the entanglement spectrum of the bipartition at low energy by that of the (1+1)D edge CFT on its entangling surface. In our case, the reduced density matrix for the A subsystem is effectively given by $\rho_{\mathrm{A}} =\rho^l_{\mathrm{CFT}}\otimes\rho^r_{\mathrm{CFT}}$,
where $\rho^l_{\mathrm{CFT}}$ (resp.~$\rho^r_{\mathrm{CFT}}$) denotes the CFT on the left (resp.~right) edge of the cylinder.

The state $\rho^l_{\mathrm{CFT}}$ at the edge of the total subsystem realizes the ground state of the edge CFT, i.e., low temperature $\beta_l \to\infty$.
Meanwhile, the state $\rho^r_{\mathrm{CFT}}$ lies at the entangling surface and corresponds to high temperature $\beta_r \to 0$ of perturbed edge CFT. 
In the following, we assume that the entanglement Hamiltonian is that of the unperturbed CFT: $\rho_{\mathrm{A};r} = e^{-\beta_r H_r}$, and check the validity of this assumption with our numerics.

% By the CFT computation, we will see that $\mathcal{T}^f_1\left(\frac{2\pi}{n}\right)$ defines an obstruction whose phase is given by $e^{\frac{2\pi i}{n}(\frac{\sigma_H}{8}-(\frac{n^2+2}{24})c_-)}\zeta^H_n$.
 
The partial rotation is evaluated in terms of the expectation values of $U_{\frac{\pi}{L},\mathrm{A}}T_{\frac{2\pi}{L},\mathrm{A}} $ acting within the CFT Hilbert space. This can be expressed as the CFT character for each edge,
\begin{align}
\begin{split}
    \mathcal{T}^f_a\left(\frac{2\pi}{n},\frac{\pi}{n}\right) &= \frac{\sum_{\alpha=\{a,a\psi\}}\mathrm{Tr}_\alpha[e^{iQ_l\frac{\pi}{n}}e^{iP_l\frac{L}{n}}e^{-\frac{\xi_l}{v}H_l}]\sum_{\alpha=\{a,a\psi\}}\mathrm{Tr}_{\alpha}[e^{iQ_r\frac{\pi}{n}}e^{iP_r\frac{L}{n}}e^{-\frac{\xi_r}{v}H_r}]}{\sum_{\alpha=\{a,a\psi\}}\mathrm{Tr}_a[e^{-\frac{\xi_l}{v}H_l}]\sum_{\alpha=\{a,a\psi\}}\mathrm{Tr}_a[e^{-\frac{\xi_r}{v}H_r}]} \\
    &= \frac{\sum_{\alpha=\{a,a\psi\}}\chi_{\alpha}(\frac{i\xi_l}{L}+\frac{1}{n};[\mathrm{AP},0],[\mathrm{AP},\frac{\pi}{n}])\sum_{\alpha=\{a,a\psi\}}\chi_{\alpha}(\frac{i\xi_r}{L}-\frac{1}{n};[\mathrm{AP},0],[\mathrm{AP},\frac{\pi}{n}])}{\sum_{\alpha=\{a,a\psi\}}\chi_{\alpha}(\frac{i\xi_l}{L};[\mathrm{AP},0],[\mathrm{AP},0])\sum_{\alpha=\{a,a\psi\}}\chi_{\alpha}(\frac{i\xi_r}{L};[\mathrm{AP},0],[\mathrm{AP},0])}
    \label{eq:rotasCFTfermion}
    \end{split}
\end{align}
where we introduced the velocity $v$, correlation length $\xi_l = v \beta_l$, $\xi_r = v \beta_r$, and the circumference of the cylinder $L$.
${P}_l$ and ${P}_r$ are translation operators on the left and right edge ${P}_l=-\frac{1}{v}H_l$, ${P}_r=\frac{1}{v}H_r$. 
$\mathrm{Tr}_\alpha[]$ denotes the trace within the sector of Hilbert space on a circle labeled by the anyon $\alpha$.
$\chi_\alpha(\tau;[s,\theta],[s',\theta])$ with $s,s'\in\{\mathrm{AP},\mathrm{P}\}$, $\theta,\theta'\in[0,\pi)$ is the CFT character that corresponds to the partition function on a torus with modular parameter $\tau$ equipped with Spin$^c$ structure. For example,
\begin{align}
    \chi_\alpha(\tau;[\mathrm{AP},\theta],[\mathrm{AP},\theta'])=\mathrm{Tr}_{\alpha,[\mathrm{AP},\theta]}[e^{iQ\theta'}e^{2\pi i \tau (L_0-\frac{c_-}{24})}]
\end{align}
where $\alpha$ labels the topological sector within the anti-periodic sector, with the $\U$ twisted boundary condition by $\theta$ in the spatial cycle. $\{L_j\}$ denotes the generator of the Virasoro algebra.
In our setup where $L \ll \xi_l$, the CFT characters for the left edge are approximated as
\begin{align}
\begin{split}
    \chi_\alpha\left(\frac{i\xi_l}{L};[\mathrm{AP},0],[\mathrm{AP},0]\right) &\approx e^{-\frac{2\pi \xi_l}{L}(h_\alpha-\frac{c_-}{24})}, \\ \chi_\alpha\left(\frac{i\xi_l}{L}+\frac{1}{n};[\mathrm{AP},0],[\mathrm{AP},\frac{\pi}{n}]\right) &\approx e^{-\frac{2\pi \xi_l}{L}(h_\alpha-\frac{c_-}{24})} e^{\frac{2\pi i}{n}(h_\alpha-\frac{c_-}{24})}.
    \end{split}
    \label{eq:approxleftedgeu1_twisted}
    \end{align}
Due to the exponential decay of the characters with respect to its spin, only the sector with the lowest spin carried by its vacuum becomes dominant in the sum over $\alpha = \{a,a\psi\}$. For simplicity of the notation, let us assume that $h_a\le h_{a\psi}$ (otherwise one can just relabel $a\rightarrow a\psi$). By plugging the above expressions into Eq.~\eqref{eq:rotasCFTfermion}, we obtain
\begin{align}
\begin{split}
    \mathcal{T}^f_a\left(\frac{2\pi}{n},\frac{\pi}{n}\right) 
    &=e^{\frac{2\pi i}{n}(h_a-\frac{c_-}{24})}\frac{\sum_{\alpha=\{a,a\psi\}}\chi_{\alpha}(\frac{i\xi_r}{L}-\frac{1}{n};[\mathrm{AP},0],[\mathrm{AP},\frac{\pi}{n}])}{\sum_{\alpha=\{a,a\psi\}}\chi_{\alpha}(\frac{i\xi_r}{L};[\mathrm{AP},0],[\mathrm{AP},0])}
    \end{split}
\end{align}
\subsubsection{The case where \texorpdfstring{$n$}{n} is odd}
From now on, let us work in cases of even/odd of $n$. When $n$ is odd, the CFT character on the right edge can be evaluated by the modular $S,T$ transformation as
\begin{align}
\begin{split}
\chi_a\left(\frac{i\xi_r}{L}-\frac{1}{n};[\mathrm{AP},0],[\mathrm{AP},\frac{\pi}{n}]\right) &=\sum_{b\in\mathcal{C}_{\mathrm{AP},\frac{\pi}{n}}} S_{ab}\chi_b\left(-\frac{1}{\frac{i\xi_r}{L}-\frac{1}{n}};[\mathrm{AP},\frac{\pi}{n}],[\mathrm{AP},0]\right) \\
&=\sum_{b\in\mathcal{C}_{\mathrm{AP},\frac{\pi}{n}}} (ST^n)_{ab}\chi_b\left(\frac{-in\frac{\xi_r}{L}}{\frac{i\xi_r}{L}-\frac{1}{n}};[\mathrm{AP},\frac{\pi}{n}],[\mathrm{AP},0]\right)
\label{eq:STcharacter}
\end{split}
\end{align}
where we used $T$ exchanges Spin$^c$ structure as
\begin{align}
T:
\begin{cases}
    ([\mathrm{AP},\theta],[\mathrm{AP},\theta'])\to ([\mathrm{AP},\theta],[\mathrm{P}+\frac{\theta+\theta'-[\theta+\theta']_{\pi}}{\pi},[\theta+\theta']_{\pi}]), \\
    ([\mathrm{AP},\theta],[\mathrm{P},\theta'])\to ([\mathrm{AP},\theta],[\mathrm{AP}+\frac{\theta+\theta'-[\theta+\theta']_{\pi}}{\pi},[\theta+\theta']_{\pi}]) \\
    \end{cases}
\end{align}
Here, $[]_{\pi}$ denotes the mod $\pi$ operation and the spin structure is acted by $\Z_2$ action as $\mathrm{AP}+1=\mathrm{P}$, $\mathrm{P}+1=\mathrm{AP}$.

The CFT Hilbert space is labeled by the boundary condition $[\text{P/AP},\theta]$ in the spatial direction, which corresponds to the $\U^f$-twisted boundary condition. According to the correspondence between the operators of chiral CFT with those of bulk topological order, the CFT states in the twisted sector $[\text{P/AP},\theta]$ are labeled by the simple objects of the bulk topological order that carries $[\text{P/AP},\theta]\in \U^f$ defect. Formally this corresponds to the simple object of the $G_b$-crossed extension followed by the modular extension of the super-modular category. That is, the states of the CFT in generic twisted sector is labeled by the objects of
\begin{align}
    \mathcal{C}_{G_f}:= \left(\bigoplus_{g_b\in G_b}\mathcal{C}_{\mathrm{AP},g_b } \right) \oplus \left(\bigoplus_{g_b\in G_b}\mathcal{C}_{\mathrm{P},g_b }\right)
\end{align}
The sum $b\in\mathcal{C}_{\mathrm{AP},\frac{\pi}{n}}$ in Eq.~\eqref{eq:STcharacter} is sum over the simple objects in the sector $\mathcal{C}_{\mathrm{AP},\frac{\pi}{n}}$.

To evaluate the action of $S,T$ modular matrices, we note that the modular transformation in the twisted Hilbert space of CFT can be read by $S,T$ matrices of $G_b$-crossed braided fusion category $\mathcal{C}_{G_f}$ that describes the topological order with global symmetry in the bulk~\cite{barkeshli2019}. In our case, the bosonic symmetry group is $G_b=\mathbb{R}/(\pi\mathbb{Z})$, and the relevant subgroup for the operation $U_{\frac{\pi}{n}}$ is the $\Z_n$ subgroup of $G_b$. Using the data of $\Z_n$-crossed braided fusion category, the $T$-matrix element is expressed as~\cite{barkeshli2019}
\begin{align}
    T^{(\mathbf{g},\mathbf{h})}_{a_{\mathbf{g}},b_{\mathbf{g}}} = e^{-\frac{2\pi i}{24}c_-}\cdot \theta_{a_{\mathrm{g}}}\cdot \eta_a(\mathbf{g},\mathbf{h})\cdot\delta_{a_{\mathbf{g}},b_{\mathbf{g}}}
\end{align}
where $\mathbf{g},\mathbf{h}\in G_b$.
Hence, the character is further rewritten as
\begin{align}
   \begin{split}
\chi_a\left(\frac{i\xi_r}{L}-\frac{1}{n};[\mathrm{AP},0],[\mathrm{AP},\frac{\pi}{n}]\right) &=\sum_{b\in\mathcal{C}_{\mathrm{AP},\frac{\pi}{n}}} (ST^n)_{ab}\chi_b\left(\frac{-in\frac{\xi_r}{L}}{\frac{i\xi_r}{L}-\frac{1}{n}};[\mathrm{AP},\frac{\pi}{n}],[\mathrm{AP},0]\right) \\
&= e^{-\frac{2\pi in}{24}c_-}\sum_{b\in\mathcal{C}_{\mathrm{AP},\frac{\pi}{n}}}  S_{ab} \theta_b^n \prod_{j=0}^{n-1} \eta_b\left(\frac{\pi}{n},\frac{j\pi}{n}\right)\cdot \chi_b\left(\frac{-in\frac{\xi_r}{L}}{\frac{i\xi_r}{L}-\frac{1}{n}};[\mathrm{AP},\frac{\pi}{n}],[\mathrm{AP},0]\right) \\
&= e^{-\frac{2\pi in}{24}c_-}\sum_{b\in\mathcal{C}_{\mathrm{AP},\frac{\pi}{n}}}\sum_{c\in\mathcal{C}_{\mathrm{AP},0} }S_{ab} \theta_b^n \prod_{j=0}^{n-1} \eta_b\left(\frac{\pi}{n},\frac{j\pi}{n}\right)\cdot S_{bc} \chi_c\left(\frac{iL}{n^2\xi_r}+\frac{1}{n};[\mathrm{AP},0],[\mathrm{AP},\frac{\pi}{n}]\right)
\label{eq:STn chia}
\end{split} 
\end{align}
The partial rotation is then expressed as
\begin{align}
    \mathcal{T}^f_a\left(\frac{2\pi}{n},\frac{\pi}{n}\right) =e^{\frac{2\pi i}{n}(h_a-\frac{c_-}{24})}\frac{e^{-\frac{2\pi in}{24}c_-}\sum_{b\in\mathcal{C}_{\mathrm{AP},\frac{\pi}{n}}}\sum_{c\in\mathcal{C}_{\mathrm{AP},0} } (S_{a,b}+S_{a\psi,b}) \theta_b^n \prod_{j=0}^{n-1} \eta_b\left(\frac{\pi}{n},\frac{j\pi}{n}\right)\cdot S_{bc}\chi_c(\frac{iL}{n^2\xi_r}+\frac{1}{n};[\mathrm{AP},0],[\mathrm{AP},\frac{\pi}{n}])}{\sum_{b\in\mathcal{C}_{\mathrm{AP},0}} (S_{a,b}+S_{a\psi,b}) \chi_b(\frac{iL}{\xi_r};[\mathrm{AP},0],[\mathrm{AP},0])}
    \label{eq:rotcharacterfermion}
\end{align}
Then, this quantity can be evaluated in the same logic as the bosonic case. For $\frac{L}{\xi_r}\gg 1$, we use the approximation that allows us to compute the CFT character only in terms of the highest weight state $\ket{h_b}$
\begin{align}
\begin{split}
    \chi_b\left(\frac{iL}{n^2\xi_r}+\frac{1}{n};[\mathrm{AP},0],[\mathrm{AP},\frac{\pi}{n}]\right) &\approx e^{\frac{2\pi i}{n}(h_b-\frac{c_-}{24})} e^{-\frac{2\pi L}{n^2\xi_r}(h_b-\frac{c_-}{24})} \\
    \chi_b\left(\frac{iL}{\xi_r};[\mathrm{AP},0],[\mathrm{AP},0]\right) &\approx e^{-\frac{2\pi L}{\xi_r}(h_b-\frac{c_-}{24})}
    \end{split}
    \label{eq:characterapproxfermion}
\end{align}
Also, due to this approximation only the state with the lowest spin $h=0$ contributes to the sum for $\frac{L}{\xi_r}\gg 1$. Hence we have
\begin{align}
\begin{split}
    \mathcal{T}^f_a\left(\frac{2\pi}{n},\frac{\pi}{n}\right) &\propto  \frac{e^{\frac{2\pi i}{n}(h_a-\frac{c_-}{24})}e^{-\frac{2\pi in}{24}c_-}\cdot\sum_{b} (S_{a,b}+S_{a\psi,b}) \theta_b^n \prod_{j=0}^{n-1} \eta_b\left(\frac{\pi}{n},\frac{j\pi}{n}\right)\cdot d_b\chi_1(\frac{iL}{n^2\xi_r}+\frac{1}{n};[\mathrm{AP},0],[\mathrm{AP},\frac{\pi}{n}])}{d_a \chi_1(\frac{iL}{\xi_r};[\mathrm{AP},0],[\mathrm{AP},0])} \\
     &\propto e^{\frac{2\pi i}{n}h_a}e^{-\frac{2\pi i}{24}(n+\frac{2}{n})c_-}\sum_{b\in\mathcal{C}_{\mathrm{AP},\frac{\pi}{n}}} \theta_b^n \prod_{j=0}^{n-1} \eta_b\left(\frac{\pi}{n},\frac{j\pi}{n}\right)\cdot d_b
    \end{split}
\end{align}
where $\mathcal{C}_{\mathrm{AP},\frac{\pi}{n}}$ is the twisted sector in $G_b$-crossed extension of super-modular category whose simple objects corresponds to the defects carrying $\pi/n\in G_b$. To relate this quantity to $\zeta^H_n$, we express the formula in terms of the anyons in the untwisted sector, instead of the twisted sector by $\frac{\pi}{n}$. This is done by using the consistency equations and the Verlinde formula of $G$-crossed braided fusion category given by~\cite{barkeshli2019}
\begin{align}
\begin{split}
    \theta_{b_{\frac{\pi}{n}}} &= \theta_{b_0}\theta_{0_{\frac{\pi}{n}}}\cdot (R^{b_0, 0_{\frac{\pi}{n}}}R^{0_{\frac{\pi}{n}}, b_0}) \\
    \eta_{b_{\frac{\pi}{n}}}\left(\frac{\pi}{n}, \frac{j\pi}{n}\right) &= \eta_{b_0}\left(\frac{\pi}{n}, \frac{j\pi}{n}\right)\eta_{0_{\frac{\pi}{n}}}\left(\frac{\pi}{n}, \frac{j\pi}{n}\right) \\
    S_{a,b_{\frac{\pi}{n}}} &= {S_{a,b_0}S_{a,0_{\frac{\pi}{n}}}}\frac{\mathcal{D}}{d_a} = \frac{1}{d_a}S_{a,b_0}(R^{a, 0_{-\frac{\pi}{n}}}R^{0_{-\frac{\pi}{n}}, a})
    \end{split}
\end{align}
where we work on the gauge with $U_{\mathbf{g}}(a,b;c)=1$. Also, one can set the gauge where $R^{a,0_{\mathbf{g}}}=R^{0_{\mathbf{g}},a}=1$~\cite{barkeshli2019}.
We then have
\begin{align}
\begin{split}
    \mathcal{T}^f_a\left(\frac{2\pi}{n},\frac{\pi}{n}\right) & \propto e^{\frac{2\pi i}{n}h_a}e^{-\frac{2\pi i}{24}(n+\frac{2}{n})c_-}\left(\theta_{0_{\frac{\pi}{n}}}^n\prod_{j=0}^{n-1}\eta_{0_{\frac{\pi}{n}}}\left(\frac{\pi}{n},\frac{j\pi}{n}\right)\right) \sum_{b\in\mathcal{C}_{\mathrm{AP},0}}    \prod_{j=0}^{n-1}\eta_b\left(\frac{\pi}{n},\frac{j\pi}{n}\right)\cdot S_{ab} d_b \theta_b^n
    \end{split}
\end{align}
where the sum is over the AP anyons in the untwisted sector. We then note that the $\U$ fractional charge is expressed as~\cite{manjunath2020FQH, Bulmash2022fermionic}
\begin{align}
    e^{i\pi Q_b} = \prod_{j=0}^{n-1}\eta_b\left(\frac{\pi}{n},\frac{j\pi}{n}\right)
\end{align}
Also, we observe the following gauge invariant quantity in the expression
\begin{align}
    \mathcal{I}_n:= \theta_{0_{\frac{\pi}{n}}}^n\prod_{j=0}^{n-1}\eta_{0_{\frac{\pi}{n}}}\left(\frac{\pi}{n},\frac{j\pi}{n}\right)
    \end{align}
We then have
\begin{align}
\begin{split}
    \mathcal{T}^f_a\left(\frac{2\pi}{n},\frac{\pi}{n}\right) & \propto e^{\frac{2\pi i}{n}h_a}e^{-\frac{2\pi i}{24}(n+\frac{2}{n})c_-}\mathcal{I}_n \sum_{b\in\mathcal{C}_{\mathrm{AP},0}}   e^{i\pi Q_b} S_{ab} d_b \theta_b^n~.
    \end{split}
\end{align}
The gauge invariant quantity $\mathcal{I}_n$ is further computed by plugging the symmetry fractionalization data into its expression, which was performed in~\cite{manjunath2020FQH}. To compute this, we can work on the specific gauge where $\theta_{0_{\bf g}}= 1$, and
\begin{align}
    \begin{split}
        \eta_{0_{\bf g}}(\bf h, \bf k) &= (F^{0_{\bf g} 0_{\bf h} 0_{\bf k}})^{-1}, \\
        F^{0_{\bf g} 0_{\bf h} 0_{\bf k}} &= \exp\left(-2\pi i (h_v+k)(z_1(z_2+z_3-[z_2+z_3]_1))\right),
    \end{split}
\end{align}
where we labeled ${\bf g} = \pi z_1, {\bf h} = \pi z_2, {\bf k} = \pi z_3$ with $z_1,z_2,z_3\in[0,1)$, and $v\in\mathcal{C}_\mathrm{P,0}$ is the vison defined as the relation $M_{b,v}= e^{i\pi Q_b}$ for all anyons $b\in\mathcal{C}_{\text{AP},0},\mathcal{C}_{\mathrm{P},0}$. The parameter $k\in\Z$ corresponds to stacking the bosonic IQH phase atop the state (here, the bosonic IQH is equivalent to eight copies of fermionic IQH state stacked with $\overline{E}_8$ state, so carries $\sigma_H=8$). The combination $h_v+k$ is identified with the electric Hall conductivity of the bosonic topological order as~\cite{manjunath2020FQH}
\begin{align}
    h_v+k = \frac{\sigma_H}{8}.
\end{align}
The quantity $\mathcal{I}_n$ is then computed as
\begin{align}
    \mathcal{I}_n = \prod_{j=1}^{n-1}\left(F^{0_{\frac{\pi}{n}} 0_{\frac{\pi}{n}} 0_{\frac{j\pi}{n}}}\right)^{-1} = \exp\left(2\pi i\frac{h_v+k}{n}\right) = e^{2\pi i \frac{\sigma_H}{8n}}.
\end{align}
The partial rotation is then simply expressed as
\begin{align}
    \mathcal{T}^f_a\left(\frac{2\pi}{n},\frac{\pi}{n}\right) & \propto e^{\frac{2\pi i}{n}h_a}e^{\frac{2\pi i}{n}(\frac{\sigma_H}{8}-(\frac{n^2+2}{24})c_-)}\sum_{b\in\mathcal{C}}   e^{i\pi Q_b} S_{ab} d_b \theta_b^n.
\end{align}
In particular, when $a=1$,
\begin{align}
    \mathcal{T}^f_1\left(\frac{2\pi}{n},\frac{\pi}{n}\right)  & \propto e^{\frac{2\pi i}{n}(\frac{\sigma_H}{8}-(\frac{n^2+2}{24})c_-)}\zeta_n^H
\end{align}
which gives an obstruction to symmetry-preserving gapped boundary.

\subsubsection{The case where \texorpdfstring{$n$}{n} is even}
A similar computation can also be done when $n$ is even. We start with the expressions in Eq.~\eqref{eq:STn chia}. One difference between odd and even $n$ is that after applying $ST^n$ transformation in the CFT character in Eq.~\eqref{eq:STn chia}, the resulting Spin$^c$ structure becomes $([\text{P},0],[\text{AP},\frac{\pi}{n}])$ instead of  $([\text{AP},0],[\text{AP},\frac{\pi}{n}])$. Then
\begin{align}
\begin{split}
    \mathcal{T}^f_a\left(\frac{2\pi}{n},\frac{\pi}{n}\right) &=e^{\frac{2\pi i}{n}(h_a-\frac{c_-}{24})}\frac{e^{-\frac{2\pi in}{24}c_-}\sum_{b,c} (S_{a,b}+S_{a\psi,b}) \theta_b^n \prod_{j=0}^{n-1} \eta_b\left(\frac{\pi}{n},\frac{j\pi}{n}\right)\cdot S_{bc}\chi_c(\frac{iL}{n^2\xi_r}+\frac{1}{n};[\mathrm{P},0],[\mathrm{AP},\frac{\pi}{n}])}{\sum_b (S_{a,b}+S_{a\psi,b}) \chi_b(\frac{iL}{\xi_r};[\mathrm{AP},0],[\mathrm{AP},0])} 
    \label{eq:rotcharacterfermioneven}
    \end{split}
\end{align}
where $b\in\mathcal{C}_{\mathrm{AP},\frac{\pi}{n}}, c\in \mathcal{C}_{\mathrm{P},0}$.  By setting the gauge where $R^{a,0_{\mathbf{g}}}=R^{0_{\mathbf{g}},a}=1$ and repeating the same logic as the case with odd $n$, we further get the expression in the leading order
\begin{align}
\begin{split}
    \mathcal{T}^f_a\left(\frac{2\pi}{n},\frac{\pi}{n}\right) &\propto e^{\frac{2\pi i}{n}h_a}e^{-\frac{2\pi i}{24}(n+\frac{1}{n})c_-}\mathcal{I}_n \sum_{b\in \mathcal{C}_{\mathrm{AP},0}}\sum_{c\in \mathcal{C}_{\mathrm{P},0}} S_{ab}\theta_b^n e^{i\pi Q_b} S_{bc}\chi_c\left(\frac{iL}{n^2\xi_r}+\frac{1}{n};[\mathrm{P},0],[\mathrm{AP},\frac{\pi}{n}]\right) \\
    &\approx e^{\frac{2\pi i}{n}h_a} e^{\frac{2\pi i}{n} h_{v'}}e^{-\frac{2\pi i}{24}(n+\frac{2}{n})c_-}\mathcal{I}_n \sum_{b\in \mathcal{C}_{\mathrm{AP},0}}\sum_{v'} S_{ab} \theta_b^n e^{i\pi Q_b} S_{bv'},
    \end{split}
\end{align}
where we sum over anyons $v'\in \mathcal{C}_{\mathrm{P},0}$ in the periodic sector with the smallest spin $h_{v'}$.
The final expression is given by
\begin{align}
    \mathcal{T}^f_a\left(\frac{2\pi}{n},\frac{\pi}{n}\right) & \propto e^{\frac{2\pi i}{n}h_a} e^{\frac{2\pi i}{n} h_{v'}}e^{\frac{2\pi i}{n}(\frac{\sigma_H}{8}-(\frac{n^2+2}{24})c_-)}\sum_{b\in \mathcal{C}_{\mathrm{AP},0}}\sum_{v'} S_{ab} \theta_b^n e^{i\pi Q_b} S_{bv'}
\end{align}

Finally, let us also compute the case with $n$ is even, and we do not perform partial U(1) transformation. In that case, we have an analogue of Eq.~\eqref{eq:rotcharacterfermion} which becomes
\begin{align}
\begin{split}
    \mathcal{T}^f_a\left(\frac{2\pi}{n},0\right) &=e^{\frac{2\pi i}{n}(h_a-\frac{c_-}{24})}\frac{e^{-\frac{2\pi in}{24}c_-}\sum_{b,c} (S_{a,b}+S_{a\psi,b}) \theta_b^n \cdot S_{bc}\chi_c(\frac{iL}{n^2\xi_r}+\frac{1}{n};[\mathrm{AP},0],[\mathrm{AP},0])}{\sum_b (S_{a,b}+S_{a\psi,b}) \chi_b(\frac{iL}{\xi_r};[\mathrm{AP},0],[\mathrm{AP},0])}.
    \label{eq:rotcharacterfermioneven_zeroU1}
    \end{split}
\end{align}
In this case, we have
\begin{align}
    \mathcal{T}^f_a\left(\frac{2\pi}{n},0\right) & \propto e^{\frac{2\pi i}{n}h_a} e^{-\frac{2\pi i}{24}(n+\frac{2}{n})c_-}\sum_{b\in \mathcal{C}_{\mathrm{AP},0}}d_b S_{ab} \theta_b^n 
\end{align}

\subsection{Bosonic case}
One can carry out a similar computation in the bosonic topological phase as well. Here we compute
\begin{align}
     \mathcal{T}^b_a\left(\frac{2\pi}{n},\frac{2\pi m}{n}\right):=\bra{\Psi_a}U_{\frac{2\pi m}{n},\mathrm{A}}T_{\frac{2\pi}{n},\mathrm{A}} \ket{\Psi_a}
\end{align}
The computation is basically the same as the fermionic case, except that the CFT character now does not depend on the spin structure of a torus. We can obtain the expression analogous to Eq.~\eqref{eq:rotcharacterfermion},
\begin{align}
    \mathcal{T}^b_a\left(\frac{2\pi}{n},\frac{2m\pi}{n}\right) =e^{\frac{2\pi i}{n}(h_a-\frac{c_-}{24})}\frac{e^{-\frac{2\pi in}{24}c_-}\sum_{b,c} S_{ab}\theta_b^n \prod_{j=0}^{n-1} \eta_b\left(\frac{2m\pi}{n},\frac{2mj\pi}{n}\right)\cdot S_{bc}\chi_c(\frac{iL}{n^2\xi_r}+\frac{1}{n};(0,\frac{2m\pi}{n}))}{\sum_b S_{ab}\chi_b(\frac{iL}{\xi_r};(0,0))}
\end{align}
where $b\in\mathcal{C}_{\frac{2m\pi}{n}}, c\in \mathcal{C}_{0}$, where $\mathcal{C}_{\mathbf{g}}$ denotes $\mathbf{g}\in\Z_n$ twisted sector of $\Z_n$-twisted category, where $\Z_n$ is understood as being embedded in $\U$. $\chi(\tau; (\mathbf{g},\mathbf{h}))$ denotes the torus with the $\Z_n$ background gauge field.
By setting the gauge where $R^{a,0_{\mathbf{g}}}=R^{0_{\mathbf{g}},a}=1$ and repeating the same logic as the fermionic case, we further get the expression in the leading order
\begin{align}
\begin{split}
    \mathcal{T}^b_a\left(\frac{2\pi}{n},\frac{\pi}{n}\right) &\propto e^{\frac{2\pi i}{n}h_a}e^{-\frac{2\pi i}{24}(n+\frac{1}{n})c_-}\mathcal{I}_{n;m} \sum_{b,c\in \mathcal{C}_{0}} S_{ab}\theta_b^n e^{2m\pi i Q_b} S_{bc}\chi_c\left(\frac{iL}{n^2\xi_r}+\frac{1}{n};\left(0,\frac{2m\pi}{n}\right)\right) \\
    &\approx e^{\frac{2\pi i}{n}h_a} e^{-\frac{2\pi i}{24}(n+\frac{2}{n})c_-}\mathcal{I}_{n;m} \sum_{b\in \mathcal{C}_{0}}S_{ab} d_b\theta_b^n e^{2m \pi i Q_b},
    \end{split}
\end{align}
where
\begin{align}
    \mathcal{I}_{n;m}:= \theta_{0_{\frac{2m\pi}{n}}}^n\prod_{j=0}^{n-1}\eta_{0_{\frac{2m\pi}{n}}}\left(\frac{2m\pi}{n},\frac{2mj\pi}{n}\right)
    \end{align}
Following~\cite{manjunath2020FQH}, we work on the specific gauge where $\theta_{0_{\bf g}}= 1$, and
\begin{align}
    \begin{split}
        \eta_{0_{\bf g}}(\bf h, \bf k) &= (F^{0_{\bf g} 0_{\bf h} 0_{\bf k}})^{-1}, \\
        F^{0_{\bf g} 0_{\bf h} 0_{\bf k}} &= \exp\left(-2\pi i (h_v+k)(z_1(z_2+z_3-[z_2+z_3]_1))\right),
    \end{split}
\end{align}
where we labeled ${\bf g} = 2\pi z_1, {\bf h} = 2\pi z_2, {\bf k} = 2\pi z_3$ with $z_1,z_2,z_3\in[0,1)$, and $v\in\mathcal{C}_0$ is the vison defined as the relation $M_{b,v}= e^{2\pi i Q_b}$ for all anyons $b\in\mathcal{C}_0$.
The parameter $k\in\Z$ corresponds to stacking the bosonic IQH phase atop the state, which carries $\sigma_H=2$. The combination $h_v+k$ is identified with the electric Hall conductivity of the bosonic topological order as  $h_v + k = \sigma_H/2$.
One can then compute $\mathcal{I}_{n;m}$ as
\begin{align}
    \mathcal{I}_{n;m}=\exp\left(2\pi i m^2\frac{h_v+k}{n}\right) = e^{2\pi i \frac{m^2}{2n}\sigma_H}.
\end{align}
We then obtain the final expression
\begin{align}
\begin{split}
    \mathcal{T}^b_a\left(\frac{2\pi}{n},\frac{2\pi m}{n}\right) & \propto e^{\frac{2\pi i}{n}(\frac{m^2}{2}\sigma_H-(\frac{n^2+2}{24})c_-)} e^{\frac{2\pi i}{n}h_a} \times\sum_{b\in \mathcal{C}} d_bS_{ab} \theta_b^n e^{2i\pi m Q_b}.
    \end{split}
\end{align}

\section{Obstructions to symmetry-preserving gapped edge state in (2+1)D fermionic Abelian FQH state}
\label{app:abelianFQE}
In this appendix, we study obstructions to symmetry-preserving gapped edge state in (2+1)D fermionic Abelian topological order with $\U$ symmetry. The super-modular tensor category for fermionic Abelian phase is expressed as $\mathcal{C} = \mathcal{C}_0\boxtimes\{1,\psi\}$ with a modular tensor category $\mathcal{C}_0$.

We find that $c_-=\sigma_H=0$, and vanishing higher central charges $\zeta_n(\mathcal{C}_0)=1$ for $\gcd(n,\frac{\nfs(\mathcal{C}_0)}{\gcd(n,\nfs(\mathcal{C}_0) )})=1$ give both necessary and sufficient conditions for symmetry-preserving gapped boundary.

First, let us check the ``sufficient'' part of the statement. $\zeta_1(\mathcal{C}_0)=1$ implies that chiral central charge carried by the bosonic theory $\mathcal{C}_0$ is 0 mod 8, so the fermionic invertible phase $\{1,\psi\}$ lies in the trivial class of sixteen fold way to realize $c_-=0$ in total. This means that once we gauge fermion parity, the resulting bosonic phase is given by untwisted $\Z_2$ gauge theory with anyons $D(\Z_2) = \{1,\psi,m,e\}$ with $m,e$ bosons, where $m$ carries the vortex of $\Z_2^f$ fermion parity.
Since $\sigma_H=0$, the vison $v$ for U(1) symmetry fractionalization must be a boson, and written as $v=v_0\times m$ with a boson $v_0\in\mathcal{C}_0$.
So far, we have seen that our fermionic phase is regarded as stacking of the following two theories:
\begin{itemize}
    \item Bosonic SET phase with U(1) symmetry, where anyons are given by a modular category $\mathcal{C}_0$, carries $c_-=0$ and U(1) symmetry fractionalization given by a vison $v_0\in \mathcal{C}_0$. Since $v_0$ is a boson, we have $\sigma_H=0$.
    \item Trivial fermionic invertible phase with $\U^f$ symmetry carrying $c_-=\sigma_H=0$.
\end{itemize}
Then, it has been shown in Ref.~\onlinecite{Kobayashi2022FQH} that the bosonic phase with U(1) symmetry satisfying $c_-=\sigma_H=0$, together with the conditions $\zeta_n(\mathcal{C}_0)=1$ for $\gcd(n,\frac{\nfs(\mathcal{C}_0)}{\gcd(n,\nfs(\mathcal{C}_0) )})=1$, admits a symmetry-preserving gapped boundary. This boundary condition for a bosonic phase $\mathcal{C}_0$ directly gives a $\U^f$ symmetry-preserving gapped boundary of our fermionic theory, so it establishes the proof of the ``sufficient'' part.

Next, let us check the ``necessary'' part of the statement. Since the requirement of $c_-=\sigma_H=0$ is obvious, we only have to show that having $\zeta_n(\mathcal{C}_0)=1$ for $\gcd(n,\frac{\nfs(\mathcal{C}_0)}{\gcd(n,\nfs(\mathcal{C}_0) )})=1$ is necessary. When the fermionic theory $\mathcal{C}$ has a gapped boundary, one can show that a bosonic theory given by minimal modular extension $\Breve{\mathcal{C}}$ of $\mathcal{C}$ has a gapped interface with the untwisted $\Z_2$ gauge theory $D(\Z_2) = \{1,\psi,m,e\}$~\cite{Kobayashi2022FQH}. Here, $\Breve{\mathcal{C}}$ physically describes a bosonic theory obtained by gauging fermion parity of the fermionic theory $\mathcal{C}$, and must have a gapped interface with the trivial fermionic phase $D(\Z_2)$ whose fermion parity has been gauged. Since $\Breve{\mathcal{C}} = \mathcal{C}_0\boxtimes D(\Z_2)$, this means $\mathcal{C}_0\boxtimes D(\Z_2)$ has a bosonic gapped interface with $D(\Z_2)$, hence $\mathcal{C}_0\boxtimes D(\Z_2)\boxtimes D(\Z_2)$ has a gapped boundary (by folding along the interface). Since $D(\Z_2)\boxtimes D(\Z_2)$ has bosonic gapped boundary, one can see that $\mathcal{C}_0$ has a bosonic gapped boundary. This is equivalent to having $\zeta_n(\mathcal{C}_0)=1$ for $\gcd(n,\frac{\nfs(\mathcal{C}_0)}{\gcd(n,\nfs(\mathcal{C}_0) )})=1$, so it establishes the ``necessary'' part of the proof.

\end{widetext}

% \onecolumngrid

% \vspace{0.3cm}

% %\input{supp_mat.tex}

% \vfill

% \onecolumngrid
% \clearpage
% \twocolumngrid

%\bibliographystyle{unsrt}

\end{document}